\documentclass[3p,times]{elsarticle}
\usepackage{graphicx}
\usepackage{amssymb}
\journal{Annals of Physics}
\begin{document}

\begin{frontmatter}
\title{An Algebraic Approach to Koopman Classical Mechanics}
\author{Peter Morgan}
\address{Physics Department, Yale University, New Haven, CT 06520, USA.}
\ead{peter.w.morgan@yale.edu}
\date{\today}

\begin{abstract}
Classical mechanics is presented here in a unary operator form, constructed using the binary multiplication and Poisson bracket operations that are given in a phase space formalism, then a Gibbs equilibrium state over this unary operator algebra is introduced, which allows the construction of a Hilbert space as a representation space of a Heisenberg algebra, giving a noncommutative operator algebraic variant of the Koopman--von Neumann approach.
In this form, the measurement theory for unary classical mechanics can be the same as and inform that for quantum mechanics, expanding classical mechanics to include noncommutative operators so that it is close to quantum mechanics, instead of attempting to squeeze quantum mechanics into a classical mechanics mold.
The measurement problem as it appears in unary classical mechanics suggests a classical signal analysis approach that can also be successfully applied to the measurement problem of quantum mechanics.
The development offers elementary mathematics that allows a formal reconciliation of ``collapse'' and ``no--collapse'' interpretations of quantum mechanics.
\end{abstract}
\begin{keyword}
Classical Mechanics, Koopman-von Neumann formalism, Quantum Mechanics
\end{keyword}
\end{frontmatter}

\newcommand\Half{{\scriptstyle\frac{\scriptstyle 1}{\raisebox{-0.4ex}{$\scriptstyle 2$}} }}
\newcommand\rmd{{\mathrm{d} }}
\newcommand\rme{{\mathrm{e} }}
\newcommand\rmj{{\mathsf{j} }}
\newcommand\Unit{{\hat{\mathbf{1}} }}
\newenvironment{myquote}[1]%
  {\list{}{\leftmargin=3em\rightmargin=#1%
           \topsep=0ex\partopsep=0ex\parsep=0ex}\item[]}%
  {\endlist}
\newenvironment{bulletpoints}%
  {\list{$\blacktriangleright$}{\leftmargin=3em%
           \topsep=0ex\partopsep=0ex\parsep=-0.3ex}\item[]}%
  {\endlist}

\vspace{-1.5ex}
\begin{bulletpoints}\large
\item The algebra of classical mechanics observables in unary form is noncommutative.
\item Measurement for this noncommutative algebra is the same as for quantum mechanics.
\item A signal analysis approach unifies the classical and quantum pictures.
\item The Liouvillian of CM is not bounded below, in contrast to the Hamiltonian of QM.
\item ``Collapse'' of the state is shown equivalent to a constraint on joint measurements.
\end{bulletpoints}

\section{Introduction}
An algebraic approach to Koopman's Hilbert space formalism for classical mechanics gives us mathematical tools that are powerful enough that we can obey Bohr's stricture that we must describe an experimental apparatus classically very closely indeed.
The noncommutativity of the transformation algebra that is introduced by the Poisson bracket is so natural for a classical physicist who fully exploits the Hilbert space tools available that not only can we classically describe the experimental apparatus, we can also describe the analysis of the experimental raw data as effectively as we can in quantum mechanics.

Many discussions now appear in the physics literature that probe the relationship between classical and quantum physics, within which Koopman's 1931 introduction of a Hilbert space formalism for classical mechanics increasingly appears.
The approach here contrasts with traditional Koopman--type approaches by being more algebra--centric, inspired by the algebraic approach to quantum mechanics\cite{LandsmanAlgQM} and to quantum field theory\cite{Haag}, so that Koopman's Hilbert space formalism for classical mechanics becomes more a classical signal analysis formalism (which introduces noncommutativity very naturally just in its free use of fourier and other integral transforms over time, which is not permitted for classical mechanics).
Experimental raw data is taken here to be a finite, lossily compressed record of an arbitrarily detailed collection of possible measurements of noisy signal voltage measurements that could have been recorded: even though we only actually perform a finite number of measurements, the set of measurements we could have performed in the past and the set of measurements we might perform in the future are both arbitrarily large.
At the level of actual signals, signal level responses to the surrounding environment do not occur as instantaneous ``collapses'': they instead transition over very short but finite times that as much depend on the material and electronic circuitry of the apparatus as on the surroundings.
The positioning of the finite number of measurements we actually perform into a continuously indexed set of possible measurements will give us a theory that discusses a field of measurements, represented by an algebra of operators, together with a relatively unstructured ``state'', which gives us information about what the results would be: it is helpful to think of the mathematics of quantum field theory as a continuously indexed field of measurements, not as a field that is measured.

The construction given here takes inspiration from many places: from quantum non--demolition measurement\cite{TsangCaves}, from the Koopman--von Neumann\cite{Koopman,Mauro,Ghose,MorganEM} and similar approaches\cite{Zalamea,BuchholzFredenhagen,Wetterich,tHooft} to classical mechanics, and from generalized probability theory\cite{Janotta}, and there is also a control systems literature that considers mixed classical and quantum models\cite{Wang,James,Gough}.
Another inspiration ---which is, however, rather flawed because it is not statistical--- is the long use of the Wigner function and other time--frequency distributions in signal analysis\cite{LeonCohen}.
This practical use of the Heisenberg algebra\cite{Kisil}, generated in the time--frequency case by $[\partial/\partial t,t]=1$, with a complex structure provided by fourier analysis, and the consequent approach to the Heisenberg uncertainty principle, has become well--known in popular science, and it may be a useful resource for some audiences, because there are several videos that present a similar idea on much--followed YouTube channels\cite{3Blue1Brown,MinutePhysics,SixtySymbols,ScienceAsylum}.
More closely, Wigner function approaches to quantum mechanics are well--known and much developed\cite{WeinbubFerry}\cite[Ch. 15]{Ballentine}, putting classical and quantum mechanics both into phase space formalisms, whereas Koopman--type approaches put classical and quantum mechanics both into Hilbert space formalisms: the two approaches have different merits, but Koopman--type approaches have thus far been much less developed.

There is also much inspiration from ``Geometric Quantization''\cite{Woodhouse}, but there is an essential contrast with that approach: quantization valiantly attempts to construct a map from the commutative algebra generated by classical phase space position and momentum observables $q$, $p$, to the noncommutative Heisenberg algebra generated by quantum observables $\hat q$, $\hat p$, but, to say it bluntly, fails.
Classical mechanics in a phase space formalism takes the observables of the theory to be functions on phase space, which does not include unary algebraic operators that can naturally be constructed using the Poisson bracket.
In phase space classical mechanics the action of the Poisson bracket as a binary operation is closed on the commutative algebra of functions on phase space ---given two functions $u$ and $v$ on phase space, $\{u,v\}$ is also a function on phase space--- but it is classically natural in a Koopman--type approach to use the unary operators that can be constructed using the Poisson bracket not only as generators of transformations.
A less constrained unary classical mechanics formalism leads to a larger, noncommutative algebra of observables, which in elementary cases amounts to allowing classical physics to use functions $u(q,p,\partial/\partial q,\partial/\partial p)$ as observables, not just functions $u(q,p)$.
A noncommutative algebra generated by $q$, $\partial/\partial q$ and by $p$, $\partial/\partial p$ is a classical mechanics that we will call CM$_+$, which, as an instance of the Heisenberg algebra, \emph{can} be mapped to two copies of quantum $\hat q$, $\hat p$.
The paper shares with Wetterich\cite{Wetterich} a concern to motivate and to justify the use of noncommutativity as a natural classical tool, but focuses on the use of the Poisson bracket and applying the methods of algebraic quantum mechanics to classical mechanics.
Given a Gibbs equilibrium state over the classical algebra we can construct a complex Hilbert space (with a complex structure provided below by the use of characteristic functions), and use that Hilbert space to model physics.

Cohn, in 1980, presented a comparable ``Operator formulation of classical mechanics''\cite{JackCohn}, from which the present approach differs in notation, by using only Gibbs equilibrium states, and by taking it to be the measurement theory that can be considered common between unary classical and quantum mechanics (despite the difference, discussed in \S\ref{Discussion}, between thermal and quantum fluctuations.)

We will begin with the simple harmonic oscillator in \S\ref{SHO}, using the binary multiplication and Poisson bracket operations to construct unary operators, and introducing a Gibbs equilibrium state, then \S\ref{SigAnalSHO} gives a signal analysis approach to the simple harmonic oscillator by introducing an explicit time parameter.
\S\ref{UnconstrainedCM} introduces more general Hamiltonians, but with the Poisson bracket structure still the same as for the simple harmonic oscillator, then \S\ref{GenPhaseSpace} discusses a more general phase space, for which both the Hamiltonian and the Poisson bracket are nontrivial.
\S\ref{GNS} gives a short discussion of the (Gelfand--Naimark--Segal) GNS--construction of a Hilbert space in a relatively elementary way, focusing on states over algebras of operators, which usefully frees us from thinking of the Hilbert space as {\sl necessarily} pre--eminent, though the familiar Hilbert space formalism will remain the first choice for practical use.

\S\ref{MeasurementsAndStates} shows how to construct states as modulations of the Gibbs equilibrium state and how to construct a noncommutative algebra of operators in a way that is classically natural, particularly from a signal analysis perspective, which can be used to model transformations that are performed either by experimental apparatus or by signal analysis algorithms whenever such sophistication is needed.
\S\ref{FigurativeDevelopment} develops and illustrates the consequences of Koopman--type mathematics for the measurement problem in \S\ref{MeasurementProblem}, suggesting a Joint Measurement Principle as a way to remove the necessity for a ``collapse'' dynamics; for Bell--type inequalities in \S\ref{BellInequalities}, applying the Joint Measurement Principle and suggesting in \S\ref{NewBell} an investigation of the variation over time of the violation of Bell--type inequalities before steady state statistics have been established; and \S\ref{Cat} applies the thinking of the preceding sections in a contrasting, figurative way to perhaps the most elementary and whimsical of examples, Schr\"odinger's cat.

\S\ref{MeasurementProblem} discusses the measurement problem in a way that is natural relative to ---but largely independent of--- the development of the mathematics of a Koopman--type approach to classical mechanics in \S\S\ref{SHO}--\ref{MeasurementsAndStates}, to some extent following the idea of Belavkin's approach\cite{Belavkin}.
After a measurement represented by an operator $\hat A$, the ``collapse'' of the state can be modeled as two steps: the first step, within the linear Hilbert space formalism, can be modeled by a projective superoperator action (known as a L\"uders transformer) on the density operator that represents a state, $\hat\rho\mapsto\hat\rho_A$; the second step can be modeled as a stochastic jump, as in \cite{Belavkin}.
Excluding the second step greatly simplifies the relevant mathematics, and, in addition, stochastic jumps are both not part of classical statistical physics at the level of the Gibbs equilibrium probability density and cannot be modeled as a linear operation.
The superoperator action of the L\"uders transformer on the state is equivalent to \emph{replacing} the collapse of the state by a superoperator action of the same L\"uders transformer on a subsequent measurement $\hat X$, a projection to the commutator subalgebra of $\hat A$, $\hat X\mapsto\hat X_A$, because of the identity, Eq. (\ref{LudersOfMeasurement}), $\mathsf{Tr}\bigl[\hat A\hat X\hat\rho_{\!A}\bigr]=\mathsf{Tr}\bigl[\hat A\hat X_{\!A}\hat\rho\bigr]$, which is exactly enough to make the subsequent measurement $\hat X_A$ jointly measurable with $\hat A$.
When joint measurements are in fact performed, they must be modeled by mutually commutative operators to ensure that in all states a joint probability density is generated by the quantum or classical Hilbert space formalism: ``collapse'' very effectively ensures, implicitly, that this is the case.
As a partial resolution of the measurement problem, excluding the final stochastic jump that has usually not been a great concern for classical physics, we can think of ``collapse'' of the state as ensuring a necessary property for joint measurements instead of as a change of the state.
The answer to the question `does this approach include ``collapse''?', which is a fundamental bellwether for interpretations, is `this is how we can answer both yes and no,' with the mathematics concerned making it possible to reconcile interpretations that differ over this hitherto rather metaphysical choice.

Koopman--type constructions can perhaps best be appreciated as offering an \emph{alternative} relationship between classical and quantum mechanics that is a significantly more unifying approach than the quantization we are used to, not as a replacement for the very effective mathematics of quantum physics.
With some subtleties, we can add a Koopman picture to the physical Schr\"odinger and Heisenberg pictures of Hilbert space mathematics.
By understanding Koopman Classical Mechanics and its relationship with Quantum Mechanics, we can hope, a little, to gain an edge in our understanding of Quantum Mechanics.

\section{The Simple Harmonic Oscillator}\label{SHO}
We will first work with a simple harmonic oscillator, for which an abstract ``position'' might perhaps be the displacement of a pendulum, but might also be a voltage or any abstract degree of freedom that is subject to a linear restorative ``force'': we will take Hamiltonian mechanics to be an abstract formalism for generating a conservative evolution equation, not a ``particle'' theory.
The position may be a point in a many--dimensional vector space, but we will work here as if it is one--dimensional, without indices (which can easily be added, however.)
A subsection, \S\ref{SigAnalSHO}, introduces a signal analysis approach that brings the simple harmonic oscillator much closer to being a field theory.

For the simple harmonic oscillator there are no constraints, so that the elementary observables of the system are straightforward functions of position and momentum, $u(q,p)$, for which we have, as well as the linear vector space structure $\lambda u(q,p)+\mu v(q,p)=[\lambda u+\mu v](q,p)$, the binary multiplication operation and the trivial binary Poisson bracket operation,
\begin{eqnarray}
    \cdot:u,v&\mapsto& [u\cdot v](q,p)=u(q,p)\cdot v(q,p),\\
    \{,\}:u,v&\mapsto& \{u,v\}(q,p)=\frac{\partial u}{\partial p}\frac{\partial v}{\partial q}
                                    -\frac{\partial u}{\partial q}\frac{\partial v}{\partial p},\label{BinaryOps}
\end{eqnarray}
with both operations being bilinear and with the latter being also a biderivation.
This structure is \emph{not} a straightforward algebra, for which there is only the linear vector space structure and \emph{one} other binary operation.
We use these binary operations to construct four linear unary operators that act on functions such as $u(q,p)$,
\begin{eqnarray}
  &&\;\hat q:u(q,p)\mapsto q\cdot u(q,p),\hspace{0.75em}
  \;\hat p:u(q,p)\mapsto p\cdot u(q,p),\nonumber\\
  &&\hat Q:u(q,p)\mapsto \{p,u\}(q,p) = \frac{\partial}{\partial q}u(q,p),\nonumber\\
  &&\hat P:u(q,p)\mapsto \{u,q\}(q,p) = \frac{\partial}{\partial p}u(q,p),
\end{eqnarray}
which can be used to construct a general unary operator as a function of $\hat q$, $\hat p$, $\hat Q$, $\hat P$, for which a product can be defined by composition, $[\hat q\hat p](w)=\hat q(\hat p(w))$, \emph{et cetera}.
For perhaps the most important example, the Hamiltonian function both becomes a multiplicative unary operator, $\hat H=\Half(\hat q^2+\hat p^2)$, and becomes the Liouvillian unary operator, $\hat L:u(q,p)\mapsto \{H,u\}(q,p)$, $\hat L=\hat p\hat Q-\hat q\hat P$.
Note carefully that this is not quantum theory, because $\hat q$ and $\hat p$ commute \emph{and} because the Liouvillian unary operator, which generates evolution over time, is not a positive operator. The algebraic structure is nonetheless closely comparable to that of quantum theory, because $\hat Q$ and $\hat P$ are both derivations, so that $[\hat Q,\hat q]=1$ and $[\hat P,\hat p]=1$, which, except for the absence of a complex structure that we will introduce below, gives two copies of the Heisenberg algebra.
Note also that we cannot present the Liouvillian operator, nor any other generators of transformations, if we do not introduce $\hat Q$ and $\hat P$, which are essential elements of the unary algebraic structure because of the Poisson bracket.
We cannot omit $\hat Q$ and $\hat P$ in a fully construed presentation of classical mechanics in a unary operator form: the functions $u(q,p)$ do not exhaust the questions that can be asked of a classical mechanical system, because, for example, it is reasonable for a classical physicist to ask whether a state is an eigenstate of the Liouvillian operator.

For a measurement theory, we look for a linear, positive, normalized \emph{state} over an algebra that is generated by $\hat q$, $\hat p$, $\hat Q$, and $\hat P$, for which we also provide the adjoint operation $\hat q^\dagger=\hat q$, $\hat p^\dagger=\hat p$, $\hat Q^\dagger=-\hat Q$, $\hat P^\dagger=-\hat P$, and, for any two unary operators, $(\hat A\hat B)^\dagger=\hat B^\dagger\hat A^\dagger$.
We interpret a state as giving the average value associated with any self--adjoint unary operator in the given state, following an algebraic quantum mechanics framework\cite{LandsmanAlgQM}, which is enough to make some kind of contact with the statistics of a collection of experimental raw data.
Other consequences can be derived, such as the association of the spectrum of an operator with the sample space of a probability density, of projection operators with a logic and with probabilities, and of average values of powers of a self--adjoint operator with higher statistical moments, \emph{et cetera}.
If we wish to emphasize the experimental interpretation of a state ---that the number it generates for a given operator is in some practical way connected to an average value of ensembles of experimental raw data--- we can call it a \emph{statistical} state.

\newcommand\kT{{\mathsf{k}_{\!B}\!\mathsf{T}}}
To construct such a state, we first note that the Gibbs equilibrium state over the phase space of the simple harmonic oscillator at finite temperature $\mathsf{T}$ and Boltzmann constant $\mathsf{k}_{\!B}$ results in average values
\begin{equation}\rho(q^{2m} p^{2n})=\int q^{2m} p^{2n}\frac{1}{2\pi\kT}\rme^{-(q^2+p^2)/2\kT}\rmd q\rmd p\\
                             =(\kT)^{m+n}\frac{(2m)!}{2^m m!}\frac{(2n)!}{2^n n!},
\end{equation}
or $\rho(q^m p^n)=0$ if either $m$ or $n$ is odd.
This can be presented in a characteristic function form as a Gaussian
\begin{equation}
  \rho(\rme^{\,\rmj\lambda q+\rmj\mu p})=\rme^{-\kT(\lambda^2+\mu^2)/2},\label{qpGibbs}
\end{equation}
which can be thought of in more elementary terms as a generating function for moments.
We can also use an inverse fourier transform to return to a probability density, which we could write informally as
\begin{equation}\label{SHOGibbsEqProb}
    P(\mathring{q},\mathring{p})=\rho\bigl(\delta(q-\mathring{q})\delta(p-\mathring{p})\bigr)
         =\frac{1}{2\pi\kT}\rme^{-(\mathring{q}^2+\mathring{p}^2)/2\kT}.
\end{equation}

The imaginary $\rmj$ has been introduced in Eq. (\ref{qpGibbs}) as an engineering convenience to allow a characteristic function to be constructed, but we will also use it as a central generator of a $*$--algebra $\mathcal{C}$ that is generated by $\hat q$, $\hat p$, $\hat Q$, $\hat P$, and $\rmj$, with adjoint $\rmj^\dagger=-\rmj$.
This introduction ---which an engineer can make carelessly for its usefulness in presenting the sine and cosine components of the fourier transform systematically even if it might give a mathematician or a philosopher pause--- also allows us to use $\rmj\hat Q$ and $\rmj\hat P$ as self-adjoint unary operators, for which measurement is relative to the fourier transform basis of improper eigenfunctions of $\rmj\hat Q$ and $\rmj\hat P$.
Other motivations for introducing a complex structure are possible, but an elementary argument can be made for considering the use of characteristic functions to be closely related to the use of complex Hilbert space methods\cite{CohenCF}.

A nonunique extension of the Gibbs equilibrium state to the algebra $\mathcal{C}$ can be constructed by using a raising and lowering operator algebra, $[a,a^\dagger]=[b,b^\dagger]=1$,
\begin{eqnarray}
  \;\hat q=(a+a^\dagger)\sqrt{\kT},&\hspace{1em}&
  \;\hat p=(b+b^\dagger)\sqrt{\kT},\cr
  \hat Q=(a-a^\dagger)/2\!\sqrt{\kT},&\hspace{1em}&
  \hat P=(b-b^\dagger)/2\!\sqrt{\kT},
\end{eqnarray}
which ensures that $[\hat Q,\hat q]=[\hat P,\hat p]=1$.
If we introduce an appropriately scaled object
\begin{equation}\hat F_{\bf f}=f_1\hat q+f_2\hat p+2\kT(f_3\rmj\hat Q+f_4\rmj\hat P),\quad
            {\bf f}\doteq(f_1,f_2,f_3,f_4),\quad \hat F_{\bf f}^\dagger=\hat F_{\bf f^*},
\end{equation}
we can construct a state that satisfies, for any unary operator $\hat A$, $\rho(a^\dagger\hat A)=\rho(b^\dagger\hat A)=\rho(\hat A a)=\rho(\hat A b)=0$ and $\rho(\hat A^\dagger)=\rho(\hat A)^*$, that is linear, $\rho(\lambda\hat A+\mu\hat B)=\lambda\rho(\hat A)+\mu\rho(\hat B)$, and that is normalized for a unit element $\Unit$, $\rho(\Unit)=1$.
We obtain, using a Baker--Campbell--Hausdorff identity, the generating function
\begin{eqnarray}
  \rho(\rme^{\,\rmj\lambda\hat F_{\bf f}})&=&\rho\left(\!\exp\!\Big[\rmj\lambda\sqrt{\kT}
             \big[(f_1+\rmj f_3)a+(f_1-\rmj f_3)a^\dagger
                      +(f_2+\rmj f_4)b+(f_2-\rmj f_4)b^\dagger\big]\Big]\right)\nonumber\\
    &=&\rme^{-\lambda^2\kT(f_1^2+f_2^2+f_3^2+f_4^2)/2},
\end{eqnarray}
which is a Gaussian characteristic function if the components of ${\bf f}=(f_1,f_2,f_3,f_4)$ are real--valued, and which is, as required, the same as in Eq. (\ref{qpGibbs}) if we set $\lambda=1$ and then $f_1=\lambda$, $f_2=\mu$, and $f_3=f_4=0$.
We define a pre--inner product
\begin{equation}({\bf f},{\bf g})\doteq\rho(\hat F_{\bf f}^\dagger\hat F_{\bf g})=\kT\left[(f_1^*+\rmj f_3^*)(g_1-\rmj g_3)+(f_2^*+\rmj f_4^*)(g_2-\rmj g_4)\right],\label{shoIP}
\end{equation}
in terms of which the commutator is
\begin{equation}[\hat F_{\bf f},\hat F_{\bf g}]=({\bf f}^*,{\bf g})-({\bf g}^*,{\bf f})
       =2\rmj\kT\left[f_3g_1-f_1g_3+f_4g_2-f_2g_4\right],
\end{equation}
so that for an arbitrary number of factors, again using the same Baker--Campbell--Hausdorff identity, we obtain for arbitrarily many unary operators $\hat F_{{\bf f}_i}$ the generating function
\vspace{0pt minus 2ex}
\begin{eqnarray}
  \rho(\rme^{\,\rmj\lambda_1\hat F_{{\bf f}_1}}\cdots\rme^{\,\rmj\lambda_n\hat F_{{\bf f}_n}})
&=&\exp\Big[-\,\Bigl(\sum_{i=1}^n\lambda_i{\bf f}_i^*,\sum_{j=1}^n\lambda_j{\bf f}_j\Bigr)/2
 -\hspace{-1.2em}\sum_{\quad 1\le i<j\le n}\hspace{-1.3em}\lambda_i\lambda_j\left[({\bf f}_i^*,{\bf f}_j)-({\bf f}_j^*,{\bf f}_i)\right]/2\Big]\label{GeneratingFunction}\\
    &&\hspace{2.05em}\mbox{noise term}\uparrow\hspace{2.9em}\mbox{incompatibility term}\uparrow\nonumber\\
&=&\exp\Big[-\sum_{i=1}^n\lambda_i^2({\bf f}_i^*,{\bf f}_i)/2
 -\hspace{-1.2em}\sum_{\quad 1\le i<j\le n}\hspace{-1.3em}\lambda_i\lambda_j({\bf f}_i^*,{\bf f}_j)\Big].
\end{eqnarray}
The first expression preserves the conceptual separation of the thermal noise term and the incompatibility term, as labeled above, whereas the second expression, which applies a straightforward algebraic simplification, is often easier to use.
From this generating function, we can use differentiation at $\lambda_i=0$ or inverse fourier transforms to construct the average value associated with any function of the $\hat F_{{\bf f}_i}$.
From Eq. (\ref{GeneratingFunction}), we can derive, for any operators $\hat A$ and $\hat B$,
\begin{equation}\rho(\hat A\rme^{\,\rmj\lambda\hat F_{\bf f}}\rme^{\,\rmj\mu\hat F_{\bf g}}\hat B)
      =\rme^{-\lambda\mu[({\bf f}^*,{\bf g})-({\bf g}^*,{\bf f})]}
               \rho(\hat A\rme^{\,\rmj\lambda\hat F_{\bf g}}\rme^{\,\rmj\mu\hat F_{\bf f}}\hat B),
\end{equation}
so that  Eq. (\ref{GeneratingFunction}) fixes the algebraic structure, for the purposes of the state, as effectively that of the Weyl--Heisenberg group.
Again using the same Baker--Campbell--Hausdorff identity, we can reduce any product of exponential factors to a single exponential of a sum,
\begin{equation}\rho(\hat A\rme^{\,\rmj\lambda\hat F_{\bf f}}\rme^{\,\rmj\mu\hat F_{\bf g}}\hat B)
     =\rme^{-\lambda\mu[({\bf f}^*,{\bf g})-({\bf g}^*,{\bf f})]/2}
              \rho(\hat A\rme^{\,\rmj\hat F_{\lambda{\bf f}+\mu{\bf g}}}\hat B),
\end{equation}
so we can in general work with terms only of the form $\rme^{\,\rmj\lambda\hat F_{\bf f}}$.
\ref{GeneratingFunctionAppendix} shows that Eq. (\ref{GeneratingFunction}) is a state in the abstract sense given in \S\ref{GNS}.

\subsection{A signal analysis approach to the simple harmonic oscillator}\label{SigAnalSHO}
Taking a signal analysis approach, we can consider the simple harmonic oscillator over time by taking ${\bf f}$ to be a function of time, ${\bf f}(t)=(f_1(t),f_2(t),f_3(t),f_4(t))$, with the algebraic  structure still given by Eq. (\ref{GeneratingFunction}), but with the time--translation invariant pre-inner product given as
\begin{equation}({\bf f},{\bf g})=\int(\tilde{\bf f}(k),\tilde{\bf g}(k))2\pi\delta(k^2-1)\frac{\rmd k}{2\pi}.
\end{equation}
The time--like evolution is restricted to fourier modes for which $k=\pm 1$, and under the fourier--mode integral we have used the pre--inner product defined above as part of the phase space construction, Eq. (\ref{shoIP}).
This gives a 1+0--dimensional random field theory, which is better thought of as a field \emph{of measurements}, indexed by functions such as ${\bf f}(t)$, instead of as a field that is measured.
The function ${\bf f}(t)$, which tells us where in time our measurement is most focused, is known as a ``window function'' in signal analysis.
As a simplest example, suppose that ${\bf f}(t)$ and ${\bf g}(t)$ have a Gaussian focus at time $t_0$ with parameter $\tau$, ${\bf f}(t)={\bf f}_0\rme^{-(t-t_0)^2/2\tau^2}/\sqrt{2\pi\tau^2}$,
${\bf g}(t)={\bf g}_0\rme^{-(t-t_0)^2/2\tau^2}/\sqrt{2\pi\tau^2}$,
so that
\begin{equation}({\bf f},{\bf g})=\int({\bf f}_0,{\bf g}_0)\rme^{-\tau^2k^2}2\pi\delta(k^2-1)\frac{\rmd k}{2\pi}
      =({\bf f}_0,{\bf g}_0)\rme^{-\tau^2}.
\end{equation}
Unsurprisingly, the thermal noise is less apparent if we average over long intervals, with large $\tau$, but we obtain the same result as for the phase space formalism if we average over short intervals, with small $\tau$.
If the oscillator interacts time--invariantly with an external time--like invariant noise, then we would expect all frequencies to be driven at different amplitudes, so that the time--translation invariant pre-inner product would be
\begin{equation}({\bf f},{\bf g})=\int(\tilde{\bf f}(k),\tilde{\bf g}(k))\tilde G(k)\frac{\rmd k}{2\pi},
\end{equation}
where $\tilde G(k)$ is the ``noise kernel'': unless a model for a real--world system includes absolutely every degree of freedom, we would have to expect that $\tilde G(k)$ will not be perfectly focused on a single fourier mode.
An analysis that freely uses mutually noncommutative integral transformations as needed ---fourier analysis!--- is a commonplace in classical signal analysis, whereas noncommutative transformations have effectively been banned from classical mechanics for a hundred years by \emph{fiat}, leaving classical mechanics as a straw man.

\begin{myquote}{0.5in}
Anticipating \S\ref{Discussion}, we can extend this structure to a Poincar\'e invariant state over a classical Klein--Gordon field just by changing the pre--inner product to
\begin{equation}
(f,g) = \hbar\!\!\int\!\!\tilde f^*(k)\,\tilde g(k)
                         2\pi\delta(k{\cdot}k-m^2)\frac{\rmd^4k}{(2\pi)^4},
\end{equation}
which can be shown to be equivalent to the quantized complex Klein--Gordon field\cite[Appendix B]{MorganEM}.
\end{myquote}

\section{The GNS--construction}\label{GNS}
We have so far obtained only a single state over the algebra of unary operators associated with the simple harmonic oscillator.
The Gelfand--Naimark--Segal construction allows us to construct a Hilbert space if we are given a single state over a $*$--algebra, which is somewhat different from a commonplace presentation of quantum theory, in which a Hilbert space is prior to the states we can construct using vectors in the Hilbert space.
With this approach, the \emph{definition} of a state replaces the introduction of the Born rule.
We here abridge the elementary account given by Haag\cite[\S III.2.2]{Haag} (see also \cite[\S 14.1.3]{Moretti}).

A \emph{state} over a $*$--algebra $\mathcal{A}$ is a linear, positive, and normalized map, $\rho:\mathcal{A}\rightarrow\mathbb{C}$, satisfying
\begin{equation}
  \rho(\lambda\hat A+\mu\hat B)\,{=}\,\lambda\rho(\hat A)+\mu\rho(\hat B),
    \qquad\rho(\hat A^\dagger\hat A)\,{\ge}\,0,\qquad\rho(\Unit)\,{=}\,1,
\end{equation}
and also commutes with the adjoint operation, $\rho(\hat A^\dagger)=\rho(\hat A)^*$.
Such a linear form defines a Hermitian scalar product for a vector space of operators $|\hat A\rangle$, $\langle\hat A|\hat B\rangle{\doteq}\rho(\hat A^\dagger\hat B){=}\bigl[\rho(\hat B^\dagger\hat A)\bigr]^{*}\!$, which is positive semi--definite, $\langle\hat A|\hat A\rangle\ge 0$, and which can be refined to a Hermitian inner product over equivalence classes, because for unary operators $\hat I$ and $\hat J$ for which $\langle\hat I|\hat I\rangle=\langle\hat J|\hat J\rangle=0$, the Schwarz inequality, $|\langle \hat A|\hat B\rangle|^2\le\langle \hat A|\hat A\rangle\langle \hat B|\hat B\rangle$, ensures that $\langle\hat A+\hat I|\hat B+\hat J\rangle=\langle\hat A|\hat B\rangle$.

We can take the vector corresponding to the unit element, $|\Unit\rangle$, to be the Gibbs equilibrium vector of the Hilbert space, $\rho(\hat A)=\langle \Unit|\hat A|\Unit\rangle$, then we can construct new vectors in the Hilbert space as $\hat A|\Unit\rangle=|\hat A\rangle$, using an arbitrary function of $\hat q$, $\hat p$, $\hat Q$, and $\hat P$, with completion in the Hilbert space norm that is given by the Hermitian inner product.
The Gibbs equilibrium vector is thus an object that we modulate by multiplication, which makes it appropriate to consider adopting $|\Unit\rangle$ as an alternative notation for the Gibbs equilibrium vector (in contrast to the usual $|0\rangle$ for the ground or vacuum state, as a zero eigenstate of all lowering operators, which greatly underplays the potency of an equilibrium or vacuum state).
New states, which should be contrasted with new vectors in the Hilbert space, can be constructed as \begin{equation}
  \rho_{\hat A}(\hat B)=\frac{\rho(\hat A^\dagger\hat B\hat A)}{\rho(\hat A^\dagger\hat A)}
            =\frac{\langle\hat A|\hat B|\hat A\rangle}{\langle\hat A|\hat A\rangle}
            =\frac{\langle \Unit|\hat A^\dagger\hat B\hat A|\Unit\rangle}{\langle \Unit|\hat A^\dagger\hat A|\Unit\rangle},
\end{equation}
or as convex sums or integrals of this construction.
For this construction, we do not necessarily need to know the structure of the $*$--algebra $\mathcal{A}$ to construct the Hilbert space, if we are given or we can in whatever way show that $\rho$ is a state, as \ref{GeneratingFunctionAppendix} shows for Eq. (\ref{GeneratingFunction}): the state fixes a representation of a class of $*$--algebras.
A state is necessary for there to be a concrete connection between the algebraic structure and statistics of a finite set of experimental raw data, so the structure of the particular representation given by a state is arguably all that is accessible.

A full account would be much more elaborate, however the bare bones of the GNS--construction are relatively elementary, with the construction of new states and new vectors being largely familiar.
The need to introduce equivalence classes and to invoke completion in the norm introduce some difficulty, but they do not have to detain us at an elementary level.

\section{A modified Hamiltonian}\label{UnconstrainedCM}
A modified Hamiltonian function determines a different dynamics than for the simple harmonic oscillator, but in this section we continue using a Poisson bracket that is globally defined as
$\{u,v\}(q,p)=\frac{\partial u}{\partial p}\frac{\partial v}{\partial q}
                                    -\frac{\partial u}{\partial q}\frac{\partial v}{\partial p}$, as for Eq. (\ref{BinaryOps}), so that we still work with representations of the Heisenberg algebra and of the Weyl--Heisenberg group that are generated by $q$, $\partial/\partial q$, and $p$, $\partial/\partial p$.
For a Hamiltonian $H(q,p)$, we replace Eq. (\ref{SHOGibbsEqProb}) by the Gibbs equilibrium probability density
\begin{equation}\label{UnconstrainedGibbsEqProb}
    \rho_H\bigl(\delta(q-\mathring{q})\delta(p-\mathring{p})\bigr)
         =\frac{1}{\mathcal{N}}\rme^{-H(\mathring{q},\mathring{p})/\kT}=P(\mathring{q},\mathring{p}),
\end{equation}
which has as generating function the fourier transform
\begin{equation}
  \rho_H(\rme^{\,\rmj\lambda q+\rmj\mu p})=\tilde P(\lambda,\mu).
\end{equation}
If we extend this state to be also over the differential operators $\partial/\partial q$ and $\partial/\partial p$, then it generates a representation of the Weyl--Heisenberg group, which therefore must be unitarily equivalent to the representation generated for the simple harmonic oscillator\cite[\S 11.5.7]{Moretti}, if the representation is irreducible.
Consequently, for some unitary operator $\hat U_{\!H}$, using the state defined for the simple harmonic oscillator by Eq. (\ref{GeneratingFunction}) (possibly for a different value of $\kT$, depending on the scale of the Hamiltonian function),
\begin{equation}
  \rho_H(\rme^{\,\rmj\lambda_1\hat F_{{\bf f}_1}}\cdots\rme^{\,\rmj\lambda_n\hat F_{{\bf f}_n}})
=\rho(\hat U_{\!H}^\dagger\rme^{\,\rmj\lambda_1\hat F_{{\bf f}_1}}\cdots\rme^{\,\rmj\lambda_n\hat F_{{\bf f}_n}}\hat U_{\!H}).\label{UnitaryEq}
\end{equation}
If the representation is reducible, which in general will be the case if the state represents a system that is in mechanical or thermodynamic contact with other systems, $\rho_H$ will be equal to a convex sum or integral of such expressions for different values of $\kT$ and for different unitaries $\hat U_{\!H}$.
For our purposes here, therefore, we will consider in Section \ref{MeasurementsAndStates} only properties of the state and Hilbert space defined by Eq. (\ref{GeneratingFunction}).

We can approach this mathematics more concretely: we can model any finite set of data by a probability density $P_H(\mathring{q},\mathring{p})$ that can be written as a convex sum of displaced Gaussian probability densities, at some temperature $\kT'$,
\begin{eqnarray}
    \rho_H\bigl(\delta(\hat q-\mathring{q})\delta(\hat p-\mathring{p})\bigr) &=&  P_H(\mathring{q},\mathring{p})
      =\!\int\!\Xi_H(\alpha,\beta)\frac{\rme^{-((\mathring{q}-\alpha)^2+(\mathring{p}-\beta)^2)/2\kT'}}{2\pi\kT'}
                                       \rmd\alpha\rmd\beta\nonumber\\
      &=&\!\int\!\Xi_H(\alpha,\beta)\rho_{[\kT']}\bigl(\delta\left(\hat q-(\mathring{q}-\alpha)\right)
                                    \delta\left(\hat p-(\mathring{p}-\beta)\right)\bigr)
                                       \rmd\alpha\rmd\beta\label{ProbabilityAsGaussian}
\end{eqnarray}
so that in characteristic function terms the convolution becomes a product,
\[\tilde P_H(\lambda,\mu)=\tilde\Xi_H(\lambda,\mu)\rme^{-\kT'(\lambda^2+\mu^2)/2}
      =\tilde\Xi_H(\lambda,\mu)\rho_{[\kT']}(\rme^{\rmj\lambda\hat q+\rmj\mu\hat p}),\qquad\tilde\Xi_H(0,0)=1,
\]
because we can always approximate that finite set of data in such a way that for a small enough choice of $\kT'$,
\[\tilde\Xi_H(\lambda,\mu)=\tilde P_H(\lambda,\mu)\rme^{+\kT'(\lambda^2+\mu^2)/2}
\]
has an inverse fourier transform.
This is a strong constraint on the fourier transform of $P_H(\mathring{q},\mathring{p})$, but any finite set of data is compatible with it.
Because $\hat Q$ and $\hat P$ generate translations, we have for any function,
$\rme^{\alpha\hat Q}F(\hat q)\rme^{-\alpha\hat Q}=F(\hat q+\alpha)$ and
$\rme^{\beta\hat P}F(\hat p)\rme^{-\beta\hat P}=F(\hat p+\beta)$, so we can construct a state $\rho_H$ as a convex sum of displaced simple harmonic oscillators,
\begin{equation}\rho_H(\rme^{\,\rmj\lambda_1\hat F_{{\bf f}_1}}\cdots\rme^{\,\rmj\lambda_n\hat F_{{\bf f}_n}})
      =\int\Xi_H(\alpha,\beta)\rho_{[\kT']}(\rme^{\alpha\hat Q+\beta\hat P}\rme^{\,\rmj\lambda_1\hat F_{{\bf f}_1}}\cdots\rme^{\,\rmj\lambda_n\hat F_{{\bf f}_n}}\rme^{-\alpha\hat Q-\beta\hat P})\rmd\alpha\rmd\beta.
\end{equation}
which is a convex sum of elementary examples of Eq.\,(\ref{UnitaryEq}).
As for state tomography in quantum mechanics, however, just knowing the probability density $P_H(\mathring{q},\mathring{p})$ does not fix the state $\rho_H$ uniquely: we can, for example, generate exactly the same probability density $P_H(\mathring{q},\mathring{p})$ using
\begin{equation}\rho_H(\rme^{\,\rmj\lambda_1\hat F_{{\bf f}_1}}\cdots\rme^{\,\rmj\lambda_n\hat F_{{\bf f}_n}})
      =\int\Xi_H(\alpha,\beta)\rho_{[\kT']}(\rme^{\alpha\hat Q+\beta\hat P+\rmj F(\alpha,\beta,\hat q,\hat p)}\rme^{\,\rmj\lambda_1\hat F_{{\bf f}_1}}\cdots\rme^{\,\rmj\lambda_n\hat F_{{\bf f}_n}}\rme^{-\alpha\hat Q-\beta\hat P-\rmj F(\alpha,\beta,\hat q,\hat p)})\rmd\alpha\rmd\beta
\end{equation}
(which is a convex sum of slightly less elementary examples of Eq.\,(\ref{UnitaryEq})), where $F(\alpha,\beta,\hat q,\hat p)^\dagger=F(\alpha,\beta,\hat q,\hat p)$, any self--adjoint function, commutes with $\hat q$ and $\hat p$, but transforms the expected measurement results for $\rmj\hat Q$ and $\rmj\hat P$.
It seems unlikely that a decomposition as a convex sum of displaced simple harmonic oscillators will be the optimal construction for use in physics, but it is perhaps the simplest example.

\section{A modified Hamiltonian {\it and} a modified Poisson bracket structure}\label{GenPhaseSpace}
If there are constraints, the elementary observables form a commutative, associative algebra $\mathcal{A}$ of functions on a phase space manifold $\mathcal{P}$, $u:\mathcal{P}\rightarrow\mathbb{R}$, with a binary multiplication operation, together with a nontrivial binary Poisson bracket operation, $\cdot:u,v\mapsto u\cdot v$, $\{,\}:u,v\mapsto \{u,v\}$, which cannot be written globally as in Eq. (\ref{BinaryOps}).
We can use the multiplication and Poisson bracket operations to construct two sets of unary operators,
\begin{eqnarray}
  \hat Y_u:\mathcal{A}\rightarrow\mathcal{A};v\mapsto \hat Y_u(v)=u\cdot v,\ \\
  \hat Z_u:\mathcal{A}\rightarrow\mathcal{A};v\mapsto \hat Z_u(v)=\{u,v\},
\end{eqnarray}
which satisfy the commutation relations
\begin{equation}
  [\hat Y_u,\hat Y_v]=0,\ \:[\hat Z_u,\hat Y_v]=\hat Y_{\{u,v\}},\ \:[\hat Z_u,\hat Z_v]=\hat Z_{\{u,v\}}.
\end{equation}
$\hat Z_u$ is called the \emph{Hamiltonian vector field} of $u$\cite[\S 3.2]{Landsman}, however this is a slightly unfortunate name insofar as only the Poisson bracket is used in its construction.
The whole algebra $\mathcal{C}$ is a semi--direct product of the $\hat Z_u$ unary operators acting adjointly on the commutative subalgebra $\mathcal{C}_Y$ that is generated by the $\hat Y_u$ unary operators, with the latter being naturally isomorphic to the multiplicative, non--Poisson bracket part of $\mathcal{A}$, $u\mapsto\hat Y_u$, $\hat Y_u(\hat Y_v(w))=\hat Y_{u\cdot v}(w)$, and $\lambda\hat Y_u(w)+\mu\hat Y_v(w)=\hat Y_{\lambda u+\mu v}(w)$, with $\hat Y_1$ as the unit element, $\hat Y_1(w)=w$.

$\mathcal{C}$ is not in general isomorphic to a Heisenberg algebra unless the classical dynamics can be written globally as in Eq. (\ref{BinaryOps}), so we cannot proceed as we did in \S\ref{UnconstrainedCM}, leading to Eq. (\ref{UnitaryEq}).
Instead, we begin by using a nontrivial Gibbs equilibrium state over $\mathcal{A}$ to construct a state over the commutative subalgebra $\mathcal{C}_Y$, normalized so that $\rho(\hat Y_1)=1$,
\begin{equation}
\rho:\mathcal{C}_Y\longrightarrow\mathbb{C};
    \rho(\hat Y_u)=\int_\mathcal{P}u(\mathbf{x})\rme^{-\,H(\mathbf{x})/\kT}\rmd\mathbf{x}
                       \hspace{0.75em}\big/\ \int_\mathcal{P}\rme^{-\,H(\mathbf{x})/\kT}\rmd\mathbf{x},
\end{equation}
with an adjoint defined as $\hat Y_u^\dagger=\hat Y_u$, $(\hat A\hat B)^\dagger=\hat B^\dagger\hat A^\dagger$, and with an imaginary $\rmj^\dagger=-\rmj$ introduced as for the simple harmonic oscillator as an efficient way to discuss the fourier sine and cosine transforms of probability distributions.
This is enough to naturally construct a state over the whole algebra $\mathcal{C}$ if we introduce the Gibbs equilibrium projection operator $\hat V$, which we can define abstractly by
$\rho(\hat A_1\hat V\cdots\hat V\hat A_n)\,{=}\,\prod_i\rho(\hat A_i)$\hspace{0.3em} (in a more elementary approach, we can define $\hat V=|\Unit\rangle\langle\Unit|$.)
A general element of the resulting algebra is $\hat A+\sum_i\hat X_i\hat V\hat A_i$,
where $\hat A,\hat A_i\in\mathcal{C}_Y$ and each $\hat X_i$ may include $\hat V$ factors.
For this general element
\begin{eqnarray}
          \rho\biggl(\!\Bigl(\hat A^\dagger+\sum\limits_{i=1}^n\hat A_i^\dagger\hat V\hat X_i^\dagger\Bigr)
                \Bigl(\hat A+\sum\limits_{j=1}^n\hat X_j\hat V\hat A_j\Bigr)\!\biggr)
   &=&\rho\biggl(\!\Bigl(\Unit\hat V\hat A^\dagger+\sum\limits_{i=1}^n\hat A_i^\dagger\hat V\hat X_i^\dagger\Bigr)
                \Bigl(\hat A\hat V\Unit+\sum\limits_{j=1}^n\hat X_j\hat V\hat A_j\Bigr)\!\biggr)\nonumber\\
     =\rho\biggl(\!\Bigl(\sum\limits_{i=0}^n\hat A_i^\dagger\hat V\hat X_i^\dagger\Bigr)
                \Bigl(\sum\limits_{j=0}^n\hat X_j\hat V\hat A_j\Bigr)\!\biggr)
                       &=&\sum\limits_{i,j}\rho(\hat A_i)^*\rho(\hat X_i^\dagger\hat X_j)\rho(\hat A_j)\ge 0,
                               \qquad\mbox{[taking $\hat X_0=\hat A$, $\hat A_0=\Unit$]}
\end{eqnarray}
so that by induction a state over $\mathcal{C}_Y$ can be extended naturally to be a state over
an algebra $\mathcal{C}_+$ of all transformations that can be constructed as a limit (in a suitable topology) of forms such as $\sum_{i,j}\alpha_{i,j}|\hat A_i\rangle\langle\hat A_j|$, with $|\hat A_i\rangle$ a countable basis of the Hilbert space $\mathcal{H}$ that is generated by the GNS--construction's action of $\mathcal{C}_Y$ on $|\Unit\rangle$.
$\mathcal{C}_+$ can include, for example, the algebra of bounded operators acting on $\mathcal{H}$, $\mathcal{B}(\mathcal{H})$.
The superoperators that can be constructed using $\mathcal{C}_+$ includes a representation of $\mathcal{C}$ as a subalgebra because $\mathcal{C}$ contains only $\mathcal{C}_Y$ and a subalgebra of adjoint actions on $\mathcal{C}_Y$.
As for unconstrained classical mechanics, we can allow all of $\mathcal{C}_+$ as observables if it is empirically necessary or useful to do so, and construct both pure and mixed states over the algebra of observables.
It can be argued from a Dutch book perspective that it is necessary to construct all such states to describe some circumstances adequately\cite{Steeger}.

Note that we have here \emph{weaponized} the Gibbs equilibrium projection operator in a very global way, because $\hat V$ does not commute with any operator in $\mathcal{C}_Y$, which is anathema to, for example, a local quantum physics perspective\cite{Haag} but is a commonplace in practical physics: whenever we use the Born rule to generate a transition probability such as $|\langle\hat B|\hat A\rangle|^2$, we implicitly use a measurement operator $|\hat B\rangle\langle\hat B|=\hat B\hat V\hat B^\dagger$ in a state with density operator $|\hat A\rangle\langle\hat A|=\hat A\hat V\hat A^\dagger$.

Constrained classical mechanics can in any case be modeled in the first instance as a limit of unconstrained classical mechanics, as in \S\ref{UnconstrainedCM}, for which the Hamiltonian function of systems that are not on the constraint surface (which is taken to be embedded in an unconstrained phase space) becomes arbitrarily large, so that in the limit the Gibbs equilibrium distribution assigns zero probability except on the constraint surface.
At the level of experimental raw data, we cannot confirm that a real experimental apparatus always lies absolutely precisely on a singular constrained surface.
We also note that quantum field theory has historically only constrained the algebra of observables to be the commutant of the symmetries of the dynamics, which mathematicians typically also restrict to $\mathcal{B}(\mathcal{H})$, which for classical mechanics we can construct as the double commutant of the Liouvillian, $\mathcal{B}(\mathcal{H})\supset\mathcal{C}=\{\hat L\}''$.
For our purposes here, therefore, we will consider below, as for unconstrained classical mechanics, only properties of the state and the Hilbert space defined by Eq. (\ref{GeneratingFunction}).

\section{Measurements and states}\label{MeasurementsAndStates}
We first focus on $\hat q$ and $\rmj\hat Q$ measurements for the simple harmonic oscillator, with $\hat p$ and $\rmj\hat P$ measurements being closely comparable.
As above, for the Gibbs equilibrium state, Eq. (\ref{SHOGibbsEqProb}),
\begin{equation}
  \rho\big(\delta(\hat q-\mathring{q})\big)=\langle \Unit|\delta(\hat q-\mathring{q})|\Unit\rangle
            =\frac{\rme^{-\mathring{q}^2/2\kT}}{\sqrt{2\pi\kT}}
\end{equation}
gives a Gaussian probability density that the position observable will be near $\mathring{q}$.
In classical mechanics, we can use the Poisson bracket to generate unitary transformations such as $\hat U=\rme^{\kappa\hat Q}$, which acts to translate the probability density, giving a different state, a \emph{modulated} form of the Gibbs equilibrium state, for which
\begin{equation}\label{displacedGibbsState}
  \langle \Unit|\hat U^\dagger\delta(\hat q-\mathring{q})\hat U|\Unit\rangle
               =\frac{\rme^{-(\mathring{q}-\kappa)^2/2\kT}}{\sqrt{2\pi\kT}},
\end{equation}
\emph{or} we can say that this is a translated measurement of the Gibbs equilibrium state.
We can further introduce convex mixtures of such states for different values of $\kappa$, such as $\rho_B\bigl(\delta(\hat q-\mathring{q})\bigr)=\sum_i\beta_i\langle \Unit|\hat U^\dagger_i\delta(\hat q-\mathring{q})\hat U_i|\Unit\rangle$, with $\sum_i\beta_i=1$, or, again, we can say that this is a more general measurement of the Gibbs equilibrium state.
We can \emph{modulate} the Gibbs equilibrium state using arbitrary polynomials of $\hat q$, taking the fourier transform with respect to $\mathring{q}$, differentiating a three factor version of the generating function Eq. (\ref{GeneratingFunction}), and taking the inverse fourier transform back to the $\mathring{q}$,
\begin{equation}
  \frac{\langle \Unit|\hat q^{n\dagger}\delta(\hat q-\mathring{q})\hat q^n|\Unit\rangle}
          {\langle \Unit|\hat q^{n\dagger}\hat q^n|\Unit\rangle}
               =\frac{2^nn!}{(2n)!}\frac{\mathring{q}^{2n}}{(\kT)^n}
                   \frac{\rme^{-\mathring{q}^2/2\kT}}{\sqrt{2\pi\kT}},
\end{equation}
so from this classical mechanics perspective we have constructed \emph{modulations} of the Gibbs equilibrium state probability density, and indeed we can by superposition and mixture construct arbitrary positive multinomial modulations of the Gibbs equilibrium state probability density, using
\begin{equation}
  \frac{\langle F(\hat q,\hat p)|\delta(\hat q-\mathring{q})\delta(\hat p-\mathring{p})|F(\hat q,\hat p)\rangle}
          {\langle F(\hat q,\hat p)|F(\hat q,\hat p)\rangle}
               =\frac{1}{\mathcal{N}}|F(\mathring{q},\mathring{p})|^2
                   \frac{\rme^{-(\mathring{q}^2+\mathring{p}^2)/2\kT}}{2\pi\kT},\qquad
\mathcal{N}\,{=}\int|F(\mathring{q},\mathring{p})|^2
                   \frac{\rme^{-(\mathring{q}^2+\mathring{p}^2)/2\kT}}{2\pi\kT}\rmd\mathring{q}\rmd\mathring{p},
\end{equation}
provided the normalization constant $\mathcal{N}$ exists.
Recalling \S\ref{UnconstrainedCM}, such states can also be thought of as the ground state probability density associated with some different Hamiltonian function.

\begin{myquote}{0.5in}
For a random field theory in operator form, such modulation of probability densities extends to there being different modulations of the Poincar\'e invariant vacuum state in different regions of space--time, a higher order analog of modulating a single--frequency carrier signal\cite{MorganEM}.
Eq. (\ref{displacedGibbsState}), for example, can be written using the pre--inner product, rather more abstractly, as
\begin{equation}
  \langle \Unit|\rme^{\,\rmj\hat F^\dagger_{\bf g}}\delta(\hat F_{\bf f}-v)\rme^{-\rmj\hat F_{\bf g}}|\Unit\rangle
               =\frac{\rme^{-(v-2\mathsf{Im}[({\bf f},{\bf g})])^2/2({\bf f},{\bf f})}}{\sqrt{2\pi({\bf f},{\bf f})}},
\end{equation}
so that more or less displacement occurs depending on the imaginary part of the pre--inner product $({\bf f},{\bf g})$.
We can understand the vacuum state as a noisy, higher order carrier, for which the noise can be considered a valuable resource for quantum computation purposes.
Note that this way of working with probabilities gives us a higher order mathematical structure than a classical field.
\end{myquote}

We can also measure components of a transformed state such as $\rho_B(\hat A)$.
For the Gibbs equilibrium state component, we can use the Gibbs equilibrium state projection operator $\hat V=|\Unit\rangle\langle \Unit|$,
$$\rho_B(\hat V)=\sum\beta_i\langle \Unit|\hat U^\dagger_i\ \hat V\ \hat U_i|\Unit\rangle,$$
or, using $|v(\hat q,\hat p)\rangle\langle v(\hat q,\hat p)|$, we can measure any other component,
$$\rho_B(|v(\hat q,\hat p)\rangle\langle v(\hat q,\hat p)|)=\sum\beta_i\langle \Unit|\hat U^\dagger_i\,|v(\hat q,\hat p)\rangle\langle v(\hat q,\hat p)|\,\hat U_i|\Unit\rangle.$$
This kind of construction is routine in quantum mechanics, but we can think of the Gibbs equilibrium state projection operator as also a classically natural measure of how much a given state is \emph{like} or \emph{unlike} the Gibbs equilibrium state, which we can use to construct comparisons with any modulated form of the Gibbs equilibrium state.
We can construct self--adjoint operators such as $|v_1\rangle\langle v_2|+|v_2\rangle\langle v_1|$, for two functions $v_1(\hat q,\hat p)$ and $v_2(\hat q,\hat p)$, as
$$|v_1\rangle\langle v_2|+|v_2\rangle\langle v_1|=\Half\big(|v_1+v_2\rangle\langle v_1+v_2|-|v_1-v_2\rangle\langle v_1-v_2|\big).$$
More generally, we can use arbitrary numbers of functions $v_i(\hat q,\hat p)$, or even more generally we can use limits in an appropriate topology of sequences of such constructions, so that using the Gibbs equilibrium state projection operator allows us to construct very general self--adjoint operators.

We can also use the well--known construction of a lowering operator $a_{\hspace{-0.05em}q}$ as an unbounded operator for which $[a_{\hspace{-0.05em}q},a^\dagger_{\hspace{-0.05em}q}]=1$ and $a_{\hspace{-0.05em}q}|\Unit\rangle=0$,
$$a_{\hspace{-0.05em}q}=\sum_{m=0}^\infty\sqrt{m+1}\,|H_m(\hat q)\rangle\langle H_{m+1}(\hat q)|,$$
where $H_m(\hat q)$ are orthonormal polynomials in $\hat q$ for which $\langle H_m(\hat q)|H_n(\hat q)\rangle=\delta_{m,n}$, constructed using $|H_0(\hat q)\rangle=|\Unit\rangle$, $\hat q|H_m(\hat q)\rangle$, and the Gram--Schmidt algorithm.
Given this kind of construction, we can think of $\hat q$ and $\hat Q$ as complementary assessments of the physical state, as systematically weighted sums and differences of how much a given state is like each of many possible states.

We can construct transformations of measurements and of states in many different ways, where different mathematical tools will correspond to different physically available objects: apparatus such as diffraction gratings, half--wave plates, \emph{et cetera}.
Transformations, which may or may not mutually commute, may be implemented in hardware, where the experimental apparatus is engineered so that a single measurement result is a function of many measurement results that could have been but are not performed, or in software, where a single measurement result is computed as a function of the records of many measurement results that have been performed, as will be the case in the Bell inequality--violating example in \S\ref{BellInequalities}.
The cycle of consecutive calibrations of mathematical models of new measurement apparatus relative to well--understood state preparations and then of mathematical models of new state preparations relative to well--understood measurement apparatus has a centuries--long history\cite{Chang}.

\subsection{Minimal and extended classical mechanics}
We can present the measurements that can be used in classical mechanics, for our purposes here, in three ways:\vspace{-0.75ex}
\begin{itemize}
\item Functions on phase space $u(q,p)$, with multiplication (na\"{\i}ve CM): this is a commutative algebra, with addition and multiplication at a point.\vspace{-0.5ex}
\item Functions on phase space $u(q,p)$, with multiplication \emph{and} the Poisson bracket: having three operations, this is not a straightforward algebra at all. However, we can convert it into a straightforward associative, non--commutative algebra of unary operators, generated by $\hat q$, $\hat p$, $\hat Q$, and $\hat P$, with $[\hat Q,\hat q]=1$ and $[\hat P,\hat p]=1$.\newline
Now there are two choices:\vspace{-0.75ex}
\begin{itemize}
\item    The Poisson bracket generated unary operators act only as transformations that leave the algebra of functions on phase space invariant.
We allow the use of $\exp(-\kappa\hat Q){\cdot}\hat q{\cdot}\exp(\kappa\hat Q){=}\hat q\,{-}\,\kappa$, but we do \emph{not} allow the use of, for example, $\exp(-\kappa\hat Q^3){\cdot}\hat q{\cdot}\exp(\kappa\hat Q^3){=}\hat q\,{-}\,3\kappa\hat Q^2$ or any other construction that gives an operator that is not a function of only $\hat q$ and $\hat p$.
The Poisson bracket binary operation phase space formalism for classical mechanics only allows this case, because $\{u,v\}(q,p)$ is indeed just another function on phase space. Call this CM$_0$.\vspace{-0.25ex}
\item    The Poisson bracket generated unary operators $\hat Q$ and $\hat P$ have the same standing as the $\hat q$ and $\hat p$ operators. The general unary operator is a function $u(\hat q, \hat p, \hat Q, \hat P)$, which is natural for a Koopman--type Hilbert space formalism for classical mechanics.
With this construction, Bell inequalities can be violated, for example, because the algebra of operators is noncommutative\cite{Landau,deMuynck,Griffiths}.
Call this CM$_+$.
\end{itemize}
\end{itemize}
Adopting CM$_+$ steps outside of the CM$_0$ that is natural for a phase space formalism for classical mechanics into what is natural for a Hilbert space formalism for classical mechanics.
A classical physicist can reasonably use CM$_+$ as a \emph{convenient}, classical tool, and cannot reasonably be stopped from using it, but does not have to use it.

\subsection{Modeling finite experimental data using matrix algebras for measurements and states}\label{FiniteModeling}
We can always present any finite amount of information about state preparations and measurements using a commutative algebra of matrices, because if we use diagonal matrices of high enough dimension $N$ we can always solve the $mn$ equations $A_{ij}=\mathsf{Tr}[\hat M_i \hat\rho_j]$ given, say, by a set of $mn$ average values of experimental raw data $\{A_{ij}, i=1..m, j=1..n\}$ for the components of $m$ diagonal measurement matrices $\hat M_i$ and the components of $n$ diagonal density matrices $\hat\rho_j$.
It is often much more \emph{convenient}, however, indeed significantly advantageous, as we know well from quantum mechanics, to solve for $\hat M_i$ and $\hat\rho_j$ as self-adjoint matrices of dimension${\ll}N$ (this process can be made to work even if we do not have averages for all $mn$ cases).
We can look for a dimension${\ll}N$ for which the information looks ``nicest'', in some sense, and we have a lot of engineering information about what numbers of dimensions work well as a first approximation for a given experiment: this process is known as quantum tomography, where ``the aim is to estimate an unknown state from outcomes of measurements performed on an ensemble of identical prepared systems''\cite{AcharyaKypraiosGuta}.
Classical mechanics can equally well adopt this kind of construction and call it a system of contextual models.

In quantum mechanics in practice, we prepare many states and measure them in many ways, however in quantum mechanics as a global metaphysics we more think of there being a single state measured in many ways, in which case in solving $A_{i}=\mathsf{Tr}[\hat M_i \hat\rho]$ there is at least one basis in which $\hat\rho$ is a diagonal matrix $\rho_j\delta_{jk}$, so that in that basis $\mathsf{Tr}[\hat M_i \hat\rho]=\sum_j M_{i,jj}\rho_j$.
If there is only one state all measurements can be taken to commute because off--diagonal entries in such a basis can be taken to be zero, so the one--state metaphysics of quantum mechanics can be taken to be the same as the one--state metaphysics of the CM$_0$ of classical physics: if we ever consider subensembles, however, the possibility of it being advantageous to use CM$_+$ re--emerges.

At a higher level, we might also find it \emph{convenient} to present information using Positive Operator Valued Measures (POVMs), using a smaller Hilbert space, even though we know by Neumark's theorem that we can always present the same information using Projection Valued Measures (PVMs) by introducing an ancilla system to construct a larger Hilbert space\cite[\S II.2.4]{BuschGrabowskiLahti}, but again we do not as classical physicists have to use this construction, it is just there for us to use if it is convenient to do so.

\section{Classical thinking about states and measurement}\label{FigurativeDevelopment}
We will develop here the idea that thinking in terms of CM$_+$ can illuminate quantum mechanics, even if we never use CM$_+$.
Many more aspects could be considered, but we will focus here on a partial solution of the measurement problem in \S\ref{MeasurementProblem}, the violation of Bell--type inequalities in \S\ref{BellInequalities}, and Schr\"odinger's cat in \S\ref{Cat}.

\subsection{The measurement problem}\label{MeasurementProblem}
We will take the measurement problem of quantum theory to concern the dynamics of ``collapse'' or ``reduction'' of the state when a measurement happens, which can be thought an unwanted contrast to the unitary dynamics that applies at all other times.
For our purposes in this subsection, we will exclude aspects of the measurement problem that concern the interpretation of probability, such as the relationship between probability and statistics and other uses of experimental raw data such as Bayesian updating, because this is also an issue for classical physics.
We will also filter the extensive literature by insisting that whatever we say about measurement must seem natural to a classical physicist who decides to adopt CM$_+$.
To that end, the principle we will apply is\vspace{1ex}
\begin{myquote}{3em}\underline{\raisebox{0ex}[0ex][0ex]{Principle (JM)}}: if we perform joint measurements that result in joint relative frequencies, then we must use mutually commutative self--adjoint operators to model those measurements.\vspace{1ex}
\end{myquote}
Principle (JM) is quite natural for a classical physicist who has previously always used CM$_0$, and, as argued in \S\ref{FiniteModeling}, we can always model finite experimental data using only commutative operators, but it is subtly not as natural in quantum mechanics, partly because although at space--like separation measurement operators are required to commute, at time--like separation measurement operators may or may not commute and we have instead become accustomed to invoking ``collapse'' of the state.
Principle (JM) is unnecessary, however, \emph{except as emphasis}, if we insist that \emph{all} measurements must be modeled by self-adjoint operators, insofar as if $\hat A^\dagger{=}\,\hat A$ and $\hat X^\dagger{=}\,\hat X$ are both self-adjoint then $\hat A\hat X$ can only be self-adjoint if $[\hat A,\hat X]\,{=}\,0$.
More explicitly, if $\hat A^\dagger{=}\,\hat A$, $\hat X^\dagger{=}\,\hat X$, and a density operator $\hat\rho^\dagger{=}\,\hat\rho$ are all mutually noncommutative, $[\hat A,\hat X]\not=0$, $[\hat A,\hat\rho]\not=0$, $[\hat X,\hat\rho]\not=0$, then
\begin{equation}
  \mathsf{Tr}\Bigl[\hat A\hat X\hat\rho\Bigr]^*=\mathsf{Tr}\Bigl[\hat\rho\hat X\hat A\Bigr]
        \not=\mathsf{Tr}\Bigl[\hat A\hat X\hat\rho\Bigr]\label{JMEq1}
\end{equation}
has an imaginary part.
Unless we ensure that either $\hat A$ or $\hat X$ always commutes with the density operator, we cannot model actual joint experimental results, which do not have an imaginary part, using operators that do not commute.

The concept of quantum non--demolition measurements, defined as mutually compatible measurements at time--like separation\cite{TsangCaves}, is well--known to quantum physics, but it is not usually insisted on when modeling joint measurements: a notable exception, however, is the \emph{Nondemolition Principle} that is introduced by Belavkin\cite{Belavkin}, which is very close in concept to Principle (JM).
\emph{We will show below how} Principle (JM) \emph{can be reconciled with the usual formalism of ``collapse''.}

In the ``collapse'' literature in quantum mechanics, the elementary linear algebraic construction asserts that after a measurement $\hat A$ that admits a discrete spectral projection $\hat A=\sum_i\alpha_i\hat P_i$, where $\hat P_i\hat P_j=\delta_{ij}\hat P_i$, $\sum_i\hat P_i=1$, for which $[\hat A,\hat P_i]=0$, a density operator $\hat\rho$ evolves instantaneously to a density operator $\hat\rho_{\!A}=\sum_i\hat P_i\hat\rho\hat P_i$, which is known as a \emph{L\"uders transformer} in the von Neumann--L\"uders measurement model\cite[\S\S II.3.2-3]{BuschGrabowskiLahti}.
\begin{myquote}{3em}This differs a little in detail from an elementary textbook description such as 
\begin{myquote}{3em}``If the particle is in a state $|\psi\rangle$, measurement of the variable (corresponding
to) $\Omega$ will yield one of the eigenvalues $\omega$ with probability
$P(\omega)\propto |\langle\omega|\psi\rangle|^2$. The state of the system will change from $|\psi\rangle$ to $|\omega\rangle$ as a result of the measurement.''\cite[\S 4.1]{Shankar},\end{myquote}
which we can model as a measurement $\hat P_i\hat A$ that not only measures $\hat A$ but also projects to (discards all but) a single eigenspace of $\hat A$ (although this does \emph{not} model a stochastic transformation, as discussed below).
[[\hspace{0.3em}{\small See also \ref{JointMeasurement} for a discussion that is less abstract than follows below.}\hspace{0.3em}]]
\end{myquote}
After a measurement $\hat A$ with this prescription, the expected value for a subsequent incommensurate measurement $\hat X$, $[\hat A,\hat X]\not=0$, will be given by
\begin{equation}\mathsf{Tr}\Bigl[\hat A\hat X\hat\rho_{\!A}\Bigr]
        =\mathsf{Tr}\Bigl[\hat A\hat X{\textstyle\sum_i}\hat P_i\hat\rho\hat P_i\Bigr]
              =\mathsf{Tr}\Bigl[{\textstyle\sum_i}\hat P_i\hat A\hat X\hat P_i\hat\rho\Bigr]
        =\mathsf{Tr}\Bigl[\hat A{\textstyle\sum_i}\hat P_i\hat X\hat P_i\hat\rho\Bigr]
        =\mathsf{Tr}\Bigl[\hat A\hat X_{\!A}\hat\rho\Bigr],\label{LudersOfMeasurement}
\end{equation}
because of the cyclic property of the trace and because $[\hat A,\hat P_i]=0$, so that the L\"uders transformer applied to the state $\hat\rho\mapsto\hat\rho_{\!A}$ is equivalent to that L\"uders transformer applied to the measurement $\hat X\mapsto\hat X_{\!A}$.
This forces mutual commutativity and allows joint measurement in a minimal way, $[\hat A,\hat X_{\!A}]=0$, $\mathsf{Tr}[\hat A\hat X_{\!A}\hat\rho]=\mathsf{Tr}[\hat X_{\!A}\hat A\hat\rho]$, \textsl{with no change of the state}.
In the quantum field theory context, where microcausality is satisfied, this identity makes it very clear that ``collapse'' only affects components of measurements that are at time--like separation from the measurement that causes the collapse: ``collapse'' is not instantaneous when considered in terms of measurements.
After joint measurement of $\hat A$ and $\hat X_{\!A}$, we can apply the same construction using the discrete spectral projection of $\hat X_{\!A}$,
$\mathsf{Tr}\bigl[\hat A\hat X_{\!A}\hat Y\hat\rho_{\!X_{\!A}}\bigr]
  =\mathsf{Tr}\bigl[\hat A\hat X_{\!A}\hat Y_{\!X_{\!A}}\hat\rho\bigr]$, \emph{et cetera}.
The discussion here focuses on joint measurement, in which context it is natural to consider the identity $\mathsf{Tr}\bigl[\hat A\hat X\hat\rho_{\!A}\bigr]=\mathsf{Tr}\bigl[\hat A\hat X_{\!A}\hat\rho\bigr]$, however we can also consider the identity $\mathsf{Tr}\bigl[\hat X\hat\rho_{\!A}\bigr]=\mathsf{Tr}\bigl[\hat X_{\!A}\hat\rho\bigr]$, which shows more concisely that a measurement $\hat X$ after the collapse of the state after a measurement $\hat A$,  $\hat\rho\mapsto\hat\rho_{\!A}$, is equivalent to a measurement $\hat X_{\!A}$ without any collapse.

The L\"uders transformer applied to subsequent measurements instead of as a collapse of the state simply \emph{enforces} Principle (JM).
For measurements that cannot be modeled by a discrete spectral projection, or when a state does not admit presentation as a density operator, we will have to ensure that whatever measurement operators we use for sequential joint measurement satisfy Principle (JM) without being able to use the L\"uders transformer as a tool.
It may in any case be best to be careful when using the L\"uders transformer, insofar as it may not be the best way to use experience to choose an operator as a first approximate model for a given measurement: the L\"uders transformer is certainly not the only way to enforce Principle (JM).
Note, however, that although idealized measurements are often modeled as having continuous sample spaces, for real experiments there is always discretization by an analog-to-digital conversion, so that real measurements can always be modeled by a discrete spectral projection.
Any observable $\hat A$ can be discretized as a binary value, using the Heaviside function, as, for a simplest example, $\hat A_{x}=\theta(x-\hat A)\,{\cdot}\,0+\theta(\hat A-x)\,{\cdot}\,1$.

The L\"uders transformer is a projection, $(\hat X_{\!A\!})_{\!A}=\hat X_{\!A}$, that maps operators to the commutant of $\hat A$, $[\hat A,\hat X_{\!A}]=0$.
If $|1\rangle$ and $|2\rangle$ are eigenvectors of $\hat A$, $\hat A|j\rangle=\alpha_{\!j}|j\rangle$,
$\hat P_i|j\rangle=\delta_{ij}|j\rangle$, we have
\begin{eqnarray*}
  \Half|1\rangle\langle 1|+\Half|2\rangle\langle 2|\hspace{0.8em}
           &\mapsto&\Half|1\rangle\langle 1|+\Half|2\rangle\langle 2|\cr
  \Half(|1\rangle+|2\rangle)(\langle 1|+\langle 2|)
           &\mapsto&\Half|1\rangle\langle 1|+\Half|2\rangle\langle 2|.
\end{eqnarray*}
The L\"uders transformer, as a linear operator, does \emph{not} model a stochastic transformation of a density operator that might be loosely written as
\begin{eqnarray}
  &&\mu|1\rangle\langle 1|+(1{-}\mu)|2\rangle\langle 2|
            \mapsto \mbox{\emph{either }}|1\rangle\langle 1|\mbox{\emph{ or }}|2\rangle\langle 2|,\nonumber\\
  &&\hspace{2em}\mbox{with probabilities $\mu$ and $1{-}\mu$ respectively,}
                                     \label{StochasticReduction}
\end{eqnarray}
which steps outside the linear algebraic representation of expected values and probability densities and, indirectly, of statistics that lossily compress the full details of the experimental raw data.
Although we \emph{can} model the discarding of states for the purposes of joint measurements by using a measurement operator such as $\hat P_i\hat A$, the L\"uders transformer for this operator is $\hat X_{\!P_{\!i\!}A}=\hat P_i\hat X\hat P_i+(1{-}\hat P_i)\hat X(1{-}\hat P_i)$, which is \emph{not} the content of Eq. (\ref{StochasticReduction}).
Such a stochastic transformation is also not modeled in the classical statistical mechanics of \S\ref{SHO}, \S\ref{UnconstrainedCM}, and \S\ref{GenPhaseSpace}, so we will not further address it here, as was declared at the beginning of this subsection, despite its obvious interest.
See Belavkin\cite{Belavkin} for one way to introduce stochastic transformations.

It might be thought that the perfect repeatability of experimental results requires collapse of the state, however the mathematics already ensures perfect correlations of repeated measurements.
If we perform a measurement modeled by $\hat A$, followed by the same measurement at a later time, modeled by $\hat B=\hat U^\dagger(t)\hat A\hat U(t)$, the joint probability density is
\begin{eqnarray}
 \rho\bigl(\delta(\hat A\,{-}\,u)\delta(\hat U(t)\hat B\hat U^\dagger(t){-}\,v)\bigr)
      &=&\rho\bigl(\delta(\hat A\,{-}\,u)\delta(\hat A\,{-}\,v)\bigr)\nonumber\\
      &=&\!\int\!\!\rho\bigl(\rme^{\,\rmj\alpha\hat A+\rmj\beta\hat A}\bigr)
                          \rme^{-\rmj\alpha u-\rmj\beta v}
                          \frac{\rmd\alpha}{2\pi}\frac{\rmd\beta}{2\pi}\nonumber\\
      &=&\!\int\!\!\rho\bigl(\rme^{\,\rmj\lambda\hat A}\bigr)
                          \rme^{-\rmj\lambda u}\delta(u\,{-}\,v)
                          \frac{\rmd\lambda}{2\pi}\nonumber\\
      &=&\rho\bigl(\delta(\hat A\,{-}\,u)\bigr)\delta(u\,{-}\,v),\label{JMEq3}
\end{eqnarray}
so that the results are \emph{perfectly} correlated without any explicit collapse mechanism being required.
This can be thought no more than an elementary, very idealized, and somewhat unrigorous version of Mott's result\cite{Mott}, that a track in a Wilson cloud chamber can be modeled by correlations already contained in a state: there is no necessity to collapse a state to model such correlations.
Conversely, if perfect correlation is \emph{not} observed, so that the results do not correspond to the mathematics above, then we have not performed the same measurement twice and we should not model the two measurements using the same operator.

The linear algebra above ---the three equations (\ref{JMEq1}), (\ref{LudersOfMeasurement}), and (\ref{JMEq3})--- and Principle (JM) \emph{can} be understood to be as natural as a way to eliminate ``collapse'' of the state as a separate dynamics in quantum mechanics as it is in classical statistical mechanics.
If we \emph{actually record} a sequence of instrument readings over time and construct joint statistics using those records, those joint statistics will be consistent with a joint probability density, but they will \emph{not} be consistent with a quasiprobability distribution that is for some joint values negative or complex.
Since for some states noncommutative operators \emph{will} generate joint distributions for real measurements that are negative or complex (which might be thought embarrassing), the operators we use to model actually recorded joint measurements had better be mutually commutative.

On this appeal to CM$_+$, there is, for example, no need for decoherence, just as there is no need for decoherence in classical probability theory when using statistics to model the throwing of a die.
A die, unlike a sphere, is typically engineered to almost always give one of six results, with care taken to use a material for which internal degrees of freedom can be ignored for most purposes: an easily malleable die could be quite different.

The idea that measurement of $\hat A$ makes it in general not possible to jointly measure $\hat A$ and $\hat X$ if $[\hat A,\hat X]\not=0$, but always possible to jointly measure $\hat A$ and $\hat X_{\!A}$ (which we might call ``$\hat X$ after $\hat A$''), is a form of contextuality, insofar as what can be measured after $\hat A$ has been measured is determined by the L\"uders transformer.
We could instead, however, jointly measure $\hat A_{\!X}$ (which we might call ``$\hat A$ modified so it does not affect $\hat X$'') with $\hat X$, because $\hat X_{\!A_{\!X}}=\hat X$.
For an elaborate discussion of joint measurement and the travails of noncontextuality, see \cite{LiangSpekkensWiseman}, however Eq. (\ref{LudersOfMeasurement}) puts the enforcement of Principle (JM) as equivalent to collapse of the state in a very compact form.

Although we \emph{can} equally apply L\"uders transformers to subsequent measurements or to the state, which we choose to do when modeling an experiment can be decided by what seems useful at the time: it is, in particular, often useful to think of a measurement as a \emph{preparation} of a state.
If microcausality is satisfied by measurements, however, it is as well to remember that applying the L\"uders transformer to a state is then to that extent not a nonlocal operation.

A comparable approach, using ``multitime correlation functions'', is suggested by \"Ottinger\cite[\S 1.2.4.1, \S 1.2.9.3]{Ottinger}, in a more elaborate formalism of quantum master equations that seeks also to model stochastic transformations of a density operator, as also does Belavkin's approach\cite{Belavkin}, which at the linear algebraic level explicitly uses quantum non-demolition measurements.

For a lucid account of the history of the measurement problem and for other recent literature, see Landsman\cite[Ch. 11]{Landsman}.
In such terms, as a counterpoint to the account above, Principle (JM) effectively formalizes the Copenhagen interpretation's requirement that an experimental apparatus in the raw and its results must be described classically, and Eq. (\ref{LudersOfMeasurement}) gives a concrete mathematical form to Bohr's doctrine of Complementarity, an idea that measurements affect the possibility of other measurements as an alternative to an idea that measurements change the state, which makes decoherence as (an example of) a mechanism to change the state unnecessary (see \cite[pp.\,4--5]{Landsman}.)
Note that we have to distinguish between variants of the Copenhagen interpretation\cite{Howard,Oldofredi}, because Bohr rejected collapse of the wave function, whereas Heisenberg, in particular, emphasized its necessity: Eq. (\ref{LudersOfMeasurement}) offers a mathematically reasoned resolution of this difference.
Landsman introduces what he calls ``\textbf{\textit{Bohrification}}, i.e., the mathematical interpretation of Bohr's classical concepts by commutative C$*$--algebras''\cite[p.\,viii]{Landsman}, which is rather close to Principle (JM) in spirit, but, without Eq. (\ref{LudersOfMeasurement}) to attribute ``collapse'' to the \emph{enforcement} of ``Bohrification'', he
\begin{myquote}{3em}{\small``describes measurement as a physical process, including the collapse that settles the outcome (as opposed to reinterpretations of the uncollapsed state, as in modal or Everettian interpretations). However, in our approach collapse takes place within unitary quantum theory.''}\cite[p.\,14]{Landsman}\end{myquote}
In contrast, we made no attempt here to discuss the final stochastic process, insofar as this problem is shared with classical statistical physics.


\subsection{The violation of Bell--type inequalities}\label{BellInequalities}
All raw data in modern experiments, whether described classically or quantum mechanically, comes into a computer along shielded signal lines attached to exotic materials that are coupled to their local surroundings and that are, furthermore, driven by support circuitry in carefully engineered ways.
What may be an elaborate hardware and software process cannot be perfectly described in a single sentence, but in outline the analog signal level is sampled and converted into binary form and the data is saved in computer storage.
There can be significant variations in this process: signal levels on many signal lines might each be stored as a 10--bit value every picosecond, say, but more typically, applying a very substantial level of compression, one or many signal levels may be analyzed for ``trigger'' conditions and information about a trigger event is stored only if a trigger condition is satisfied.

At the level of the signal picosecond by picosecond, there is in a sense no collapse, but if material properties and electronic circuitry are engineered to give a signal that satisfies a trigger condition only occasionally, then we \emph{can} think of that moment as a collapse, but we can \emph{also} think of it as very like the classical moment when we decide that a die has finally settled after it has landed and rolled a few times, before we pick up the die to throw it again.

Crucially in what follows, an empiricist approach should not too quickly assume that the satisfaction of signal level trigger conditions is caused by a ``particle'' or any other isolated system: from a classical signal analysis perspective, the signal levels are more appropriately associated with whatever locally surrounds the exotic materials and circuitry that directly drive the signal levels, with the whole experimental apparatus being driven in turn by other exotic materials and circuitry that are relatively remote, tens or thousands of meters away.
Certainly an operator algebra may include idealized operators that have a discrete spectrum (or, classically, a discrete sample space), as well as operators that more realistically model the finite widths of real spectra, but we should not ---or, again, not too quickly--- assume that a discrete spectrum implies that there are point particles that cause that discreteness.

\begin{figure}[b]
\centerline{\includegraphics[scale=1.0]{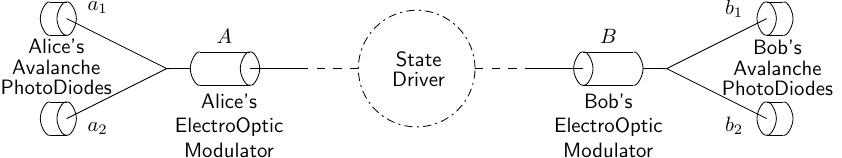}}
\caption{A schematic diagram of Weihs' experiment to violate a Bell--type inequality. Six signals from the apparatus are recorded in a lossily compressed form: $a_1$, $a_2$, $A$, $b_1$, $b_2$, and $B$.\label{WeihsDiagram}}
\end{figure}

For a review of the violation of Bell--type inequalities, see \cite{BellReview}.
For a specific experiment that violates a Bell--type inequality, we consider the measurements that were performed by Weihs\cite{Weihs,WeihsThesis}, for which see Fig. \ref{WeihsDiagram}, which we take to be a compressed description of six measured voltages, $q_1(t), q_2(t), q_3(t), q_4(t), q_5(t), q_6(t)$, on signal lines attached to four Avalanche PhotoDiodes (APDs), two for Alice, $a_1{\,=\,}q_1$, $a_2{\,=\,}q_2$, and two for Bob, $b_1{\,=\,}q_4$, $b_2{\,=\,}q_5$, and to two ElectroOptic Modulators (EOMs), $A{\,=\,}q_3$, $B{\,=\,}q_6$.
See \cite[page 60]{WeihsThesis} for a beautifully clear schematic of the experiment.
All six of these voltages are the output of electronic systems that both have complex dynamics of their own and are externally driven in complex ways, enough that we cannot measure the associated momenta.
The trigger conditions, however, implicitly use crude assessments of the time derivatives of the signals over device appropriate time scales.
The APD signals are mostly near zero voltage, but, at random intervals, on average once every $\sim$100 $\mu$s during the Weihs experiment, each APD signal becomes and stays near a larger voltage, which we will call ``1'', for $\sim$1 $\mu$s (the ``dead time''), so that each APD signal level averaged over long periods is $\sim$0.01.
Considered over long periods each APD is in a time--translation invariant thermodynamic equilibrium, whereas over short periods each APD is in a thermodynamically metastable state that is engineered to be easily disturbed.
The APDs, and the state of the whole experimental apparatus, are driven by a single light source, through a relatively exotic crystal, in a way that results in elaborate correlations between the signals $a_1$, $a_2$, $b_1$, and $b_2$.
The two EOMs are driven by external voltages that are as statistically independent as possible, so that the signal level at each EOM might or might not change between ``0'' and ``1'' every $\sim$0.1 $\mu$s.

The raw signal level data, if it were recorded as six voltages averaged over a picosecond timescale and digitized to 10--bit accuracy, would be 8 Terabytes per second, so it was lossily compressed in hardware by recording the time at which each APD signal changes from 0 to 1, as a 64--bit number, with accuracy $\sim$0.5 ns, together with the local EOM signal ($A$ for $a_1$ and $a_2$, $B$ for $b_1$ and $b_2$), \emph{provided that} the local EOM signal is not at the same time changing between the values 0 and 1. The compressed data rate was therefore $\sim$100 kilobytes per second per APD.

This is all that was done during each experimental run, which was followed much later by signal analysis in which signal transition times were compared and pairs of transition times that were close to coincident (within a few nanoseconds) were collated into 16 categories, for events on the $a_1$ or $a_2$ signal lines and for the EOM signal line $A$ either 0 or 1, and similarly for Bob's signal lines.
For example, for data from the experimental run \emph{longdist35}, which can be obtained on reasonable request from Gregor Weihs, we can construct a table such as Table \ref{WeihsResults}
\begin{table}
\centerline{\begin{tabular}{|c|c||c|c|c|c|}
\hline
\multicolumn{2}{|c||}{}
&\multicolumn{2}{|c|}{$A=0$}&\multicolumn{2}{|c|}{\rule{0ex}{2.5ex}$A=1$}\\
\cline{3-6}
\multicolumn{2}{|c||}{Bob\raisebox{1ex}[0pt]{\hspace{0.5em}$\diagdown$}
                        \raisebox{2.5ex}[0pt]{Alice}}&$a_1$&$a_2$&$a_1$&$a_2$\\
\hline\multicolumn{2}{|c||}{\vspace{-2.65ex}}&{}&{}&{}&{}\\\hline
{} & $b_1$ &320 & 1780 & 1675 & 193\\
\cline{2-6}
\raisebox{1.5ex}[0pt]{$B=0$} & $b_2$ &2006 & 364 & 300 & 1212\\
\hline
{} & $b_1$ & 439 & 1741 & 293 & 1200\\
\cline{2-6}
\raisebox{1.5ex}[0pt]{$B=1$} & $b_2$ &1658 & 374 & 1463 & 181\\
\hline
\end{tabular}}
\caption{Counts for approximately coincident events on APD signal lines $a_1$, $a_2$, $b_1$, and $b_2$, conditioned by EOM signals $A$ and $B$ at the time.\label{WeihsResults}}
\end{table}%
(with general properties as shown, but with the precise numbers depending on details of the algorithm and its parameters).
The 16 observables in Table \ref{WeihsResults} are in four groups, corresponding to events post--selected according to the EOM settings, for which we compute relative frequencies,
\begin{eqnarray}
  &&E_{00}=\frac{320-2006-1780+364}{320+2006+1780+364} = -0.694;\nonumber\\
  &&E_{10}=\frac{1675-300-193+1212}{1675+300+193+1212} = 0.708;\nonumber\\
  &&E_{01}=\frac{439-1658-1741+374}{439+1658+1741+374} = -0.614;\nonumber\\
  &&E_{11}=\frac{293-1463-1200+181}{293+1463+1200+181} = -0.698;\nonumber\\
  &&\hspace{5em}|E_{00}+E_{01}+E_{11}-E_{10}|=2.714.\label{BellExpression}
\end{eqnarray}
The last expression exhibits the violation of a Bell--type inequality, which shows that the observables concerned \emph{cannot} be elements of a commutative algebra\cite{Landau,deMuynck,Griffiths}.
Ordinarily, this would be a death knell for classical mechanics, but it is not for CM$_+$.

We can apply Principle (JM) to this description of the Weihs experiment: measurements of the six $q_i(t)$ over time are joint measurements, as is the lossily compressed data that is actually stored, so they should be modeled by mutually commuting operators (even though the experiment as performed only records one instance at each time and place.)
The signal analysis algorithm, however, constructs statistics for events on the $a_1$ or $a_2$ signal lines and for the EOM signal line $A$ being \emph{either} 0 \emph{or} 1: we \emph{cannot} make joint measurements of the occurrence of events on the $a_1$ signal line when the EOM signal line A is \emph{both} 0 \emph{and} 1, so such statistics may well have to be modeled by noncommuting operators ---which, again, \emph{can} be done within CM$_+$.

A typical quantum theoretical model for this experiment introduces two 2--dimensional Hilbert spaces, $\mathcal{H}_{\mathrm{Alice}}$, spanned by
$|H_{\mathrm{Alice}}\rangle$ and $|V_{\mathrm{Alice}}\rangle$,
 and $\mathcal{H}_{\mathrm{Bob}}$, spanned by
$|H_{\mathrm{Bob}}\rangle$ and $|V_{\mathrm{Bob}}\rangle$, and the 4--dimensional tensor product
$\mathcal{H}_{\mathrm{Alice}}\otimes\mathcal{H}_{\mathrm{Bob}}$.
Relative frequencies of $a_1$ and $a_2$ events can be represented by one pair of orthogonal projection operators $\hat A_1$ and $\hat A_2$ acting on $\mathcal{H}_{\mathrm{Alice}}$ for $A\,{=}\,0$ and by a different pair of orthogonal projection operators $\hat A_1'$ and $\hat A_2'$ acting on $\mathcal{H}_{\mathrm{Alice}}$ for $A\,{=}\,1$, and similarly for relative frequencies of $b_1$ and $b_2$ events for $B\,{=}\,0$ and $B\,{=}\,1$.
$\hat A_1$, $\hat A_2$, $\hat A_1'$, and $\hat A_2'$ generate a noncommutative algebra of operators $\mathcal{A}$;
$\hat B_1$, $\hat B_2$, $\hat B_1'$, and $\hat B_2'$ generate a noncommutative algebra of operators $\mathcal{B}$, all of which commute with all of $\mathcal{A}$; and together they generate an algebra of operators $\mathcal{A}\vee\mathcal{B}$.
Following Landau's derivation\cite{Landau} (but see also \cite[\S I.A]{BellReview}), define
$\hat\mathsf{a}\,{=}\,\hat A_1{-}\hat A_2$, $\hat\mathsf{b}\,{=}\,\hat B_1{-}\hat B_2$,
$\hat\mathsf{a}'\,{=}\,\hat A_1'{-}\hat A_2'$, and $\hat\mathsf{b}'\,{=}\,\hat B_1'{-}\hat B_2'$, for which $\hat\mathsf{a}^2\,{=}\,\hat\mathsf{b}^2\,{=}\,\hat\mathsf{a}'{}^2\,{=}\,\hat\mathsf{b}'{}^2\,{=}\,1$, and define
$$\hat\mathsf{C}=\hat\mathsf{a}\hat\mathsf{b}+\hat\mathsf{a}\hat\mathsf{b}'
                          +\hat\mathsf{a}'\hat\mathsf{b}'-\hat\mathsf{a}'\hat\mathsf{b},$$
for which we find eight terms in $\hat\mathsf{C}^2$ cancel, leaving $\hat\mathsf{C}^2\,{=}\,4+[\hat\mathsf{a},\hat\mathsf{a}'][\hat\mathsf{b},\hat\mathsf{b}']$.
For CM$_0$, we would obtain, in a state $\rho$, because both commutators must be zero, $|\rho(\hat\mathsf{C})|^2\le \rho(\hat\mathsf{C}^2)=4$, whereas for CM$_+$ and QM we obtain, because for the spectral norm we have $\|\hat\mathsf{a}\|\,{=}\,\|\hat\mathsf{b}\|\,{=}\,\|\hat\mathsf{a}'\|\,{=}\,\|\hat\mathsf{b}'\|\,{=}\,1$ and hence $\|[\hat\mathsf{a},\hat\mathsf{a}'][\hat\mathsf{b},\hat\mathsf{b}']\|\le 4$, $\|\hat\mathsf{C}^2\|\le 8$,
$|\rho(\hat\mathsf{C})|^2\le \rho(\hat\mathsf{C}^2)\le 8$.
An extremal $4{\times}4$ matrix model of the above algebraic structure and a state for which $|\rho(\hat C)|\,{=}\,2\sqrt{2}\,{>}\,2$ is given in \ref{MatrixRep}.
In Eq. (\ref{BellExpression}) above,
$E_{00}\,{\sim}\,\rho(\hat\mathsf{a}\hat\mathsf{b})$,
$E_{10}\,{\sim}\,\rho(\hat\mathsf{a}'\hat\mathsf{b})$,
$E_{01}\,{\sim}\,\rho(\hat\mathsf{a}\hat\mathsf{b}')$, and
$E_{11}\,{\sim}\,\rho(\hat\mathsf{a}'\hat\mathsf{b}')$.
Other Bell--type inequalities that depend on whether the algebra of operators that are allowed in models is commutative or noncommutative can be derived.

The above derivation is independent of locality, however a state over $\mathcal{A}\vee\mathcal{B}$ that generates relative frequencies close to those in Table \ref{WeihsResults} is nonlocal in the sense that it is not a product of a state over $\mathcal{A}$ and another state over $\mathcal{B}$.
As superficially perplexing as this is usually thought to be, a nonlocality that is determined by the boundary conditions of the experimental apparatus is \emph{expected} for a classical time--translation invariant equilibrium state, with equilibrium being established more or less quickly depending on the dynamics at smaller scale.
Such a 4--dimensional Hilbert space model derives from an infinite--dimensional Hilbert space that models the electromagnetic field state of the whole of the experimental apparatus, but we can as much take that Hilbert space and measurement operators that act on it to be generated by a random field\cite{MorganEM} as we usually take it to be generated by a quantum field; Bell inequalities for random fields are also discussed in \cite{MorganBellForRandom}.
Note that taking APD events to be a result of coupling of the APDs to the electromagnetic field is an alternative to taking APD events to be measurements of particle properties.

\subsubsection{An experimental proposal}\label{NewBell}
For the experiment above, the statistics of events on the signal lines are time--translation invariant, in that we turn on the power to a laser that drives the state, wait some reasonable length of time, then collect data, but if we waited for a slightly longer or shorter length of time we would obtain a very similar violation of a Bell--type inequality.
A signal analysis approach to such data suggests that it would be both theoretically and technologically useful to characterize how the violation evolves over time immediately after the power to the laser is turned on, because it is surely not guaranteed \emph{a priori} that the numbers of coincident events and the violation of the Bell--type inequality will increase precisely as fast as the number of events on each signal line separately.
Indeed, insofar as the violation of Bell--type inequalities in such experiments is a time--translation invariant equilibrium condition, we would more expect the approach to an equilibrium nonlocal state that results in the violation of Bell--type inequalities might be slower than the approach to a local equilibrium of events on each signal line: it would be interesting to know by precisely how much, or whether it is in fact just as fast.
It may well be that for some applications we cannot leave the power to the laser permanently on, in which case how the numbers of coincident events and the violation of Bell--type inequalities change over time after power is supplied may also be technologically important.
It may also be that the numbers of coincident events and the violation of Bell--type inequalities will increase at different rates when the separation between Alice and Bob is increased or when different optical components are used, or subtle universalities may emerge even for apparently very different experiments.
It would seem significant, for example, if it takes time equivalent to some large multiple of the length of the whole apparatus before violation of Bell--type inequalities is established.

To investigate such variation over time, we can turn on the power to the laser, wait a millisecond (or more or less, if we discover this is shorter or longer than necessary), turn off the power and wait until the event rate decreases to the dark rate, and repeat this process until we have enough statistics.
From the resulting data, we can compute counts as above for Table \ref{WeihsResults} for events in every 0.1$\mu$s time slice, say, after the power is turned on, and plot the increase of single events on each APD signal line, the numbers of coincident events, and the violation of the Bell--type inequality over time after the power is turned on.
If we wish to understand the dynamics of such experiments, instead of essentially static statistics, we should investigate the variation of coincidences and of the violation of Bell--type inequalities over time.
An alternative computation, which can be performed using the same data, would consider how the violation of Bell--type inequalties is different for an ensemble of the very first coincident pairs after each time the power is turned on, for an ensemble of the second coincident pairs, and so on, independent of the timings of the coincident pairs.

Consideration of electromagnetic field observables as well as of avalanche event timings in APDs also suggests that correlations between the detailed APD signal voltages averaged over, say, nanosecond or picosecond periods (conditioned both on the EOM signals $A$ and $B$ and on which of the four APD signals $a_1$, $a_2$, $b_1$, and $b_2$ are currently in their avalanche state) should be examined for whether correlations of avalanche timings and the associated violation of Bell--type inequalities can also be detected in other, less compressed experimental raw data\cite[\S VII.C.2]{BellReview}.
The APDs are coupled to the electromagnetic field that is driven by the electrically active components of the experiment and modified and contained by exotic materials, wave guides, and other electrically inactive components of the experiment, so that there should be \emph{some} correlations between the detailed APD signal voltages and the electromagnetic field.
It will be interesting to know, however, whether APD events are effectively ``rogue wave''--type events, for which precursor conditions that allow prediction are typically very difficult to identify.

\subsection{Schr\"odinger's cat: what's the state?}\label{Cat}
Suppose a quantum physicist prepares a box and tells a classical physicist that in the box there is a cat that is in a superposition of being alive and being dead.
It's a little whimsical to ask, as whimsical perhaps as thinking about cats in the context of quantum mechanics is always bound to be, but how can the classical physicist be sure whether the quantum physicist is telling the truth?
Most often there is a song and a dance about decay of a nucleus that is initially assumed to be in a pure state and about unitary transformation of the nucleus and cat\cite[\S I A]{Frowis}, but idealized claims about a state preparation have to be backed up with experimental verification that all possible effects of other degrees of freedom have been completely enough eliminated\cite{Serafini}.
Is the state a superposition or a mixture?
We focus here only on the extent to which the answer depends on whether we allow a noncommutative algebra of operators as models for measurements.

Both classically and quantum mechanically, suppose that when we open the box and we measure whether the cat is alive, using a projection operator that we can present as $\hat A=\left(\!\!\begin{array}{cc}1 & 0\\0 & 0\end{array}\!\!\right)$, acting on a Hilbert space that is, as in \S\ref{BellInequalities}, spanned by two vectors, $|\mathrm{Alive}\rangle$ and $|\mathrm{Dead}\rangle$, but that derives from a \emph{much} higher--dimensional Hilbert space, in either CM$_0$, CM$_+$, or QM.
Then through some experimental sleight of hand ---an ensemble of cats in boxes--- we obtain a probability $\alpha$ that the cat is alive.
We can represent that result using a density matrix
$\left(\!\!\!\begin{array}{cc}\alpha \!&\! \beta\\\beta^* \!&\! 1{-}\alpha\end{array}\!\!\!\right)$, which is consistent with either a mixed state such as
$\hat M_\alpha\,{=}\,\left(\!\!\!\begin{array}{cc}\alpha \!&\! 0\\0 \!&\! 1{-}\alpha\end{array}\!\!\!\right)$
or a pure state such as $\hat S_\alpha\,{=}\,\left(\!\begin{array}{cc}\alpha\!\! & \!\!\sqrt{\alpha(1{-}\alpha)}\!\\ \!\!\sqrt{\alpha(1{-}\alpha)}\!\! & \!\!1{-}\alpha\end{array}\!\right)$, for all of which we obtain precisely the same probability $\alpha$ that the cat is alive.
To tell whether the quantum physicist is telling the truth, the classical physicist \emph{must} use other observables, such as what could be called the Lewis Carroll operators, $\hat C_1=\left(\!\!\begin{array}{cc}0 & 1\\1 & 0\end{array}\!\!\right)$ and  $\hat C_2=\left(\!\!\begin{array}{cc}0 & \rmj\\-\rmj & 0\end{array}\!\!\right)$, each of which, in slightly different ways, takes a live cat and kills it and takes a dead cat and resuscitates it: a little strange and very difficult to implement, but comprehensible to a Victorian mathematician and in CM$_{+}$.
With these operators as well as $\hat A$, the classical physicist can determine $\beta$, which likely eliminates both $\beta=0$ and $\beta=\sqrt{\alpha(1{-}\alpha)}$ as possibilities, which \emph{cannot} be done if the classical physicist only uses operators that are compatible with $\hat A$.

According to the usual account, a classical physicist's measurements are always mutually commutative, which can even be held up as the fundamental difference between classical and quantum.
In that case, the classical physicist cannot tell whether the alleged preparation is what the quantum physicist says it is or not.
If the classical physicist accepts that all their measurements are and must be mutually commutative, they can reasonably say, ``Huh, it's just a mixture, which I understand well enough, you're just muddying perfectly clear waters by saying that it's a superposition''.
In fact, however, the Lewis Carroll operators are classically well--enough--defined.
If the classical physicist allows themselves to use the Lewis Carroll and similar operators, then they can tell whether the state is a pure state, and they can confirm all the quantum physicist's claims, but with that expansion of what a classical physicist can do, to CM$_+$ instead of CM$_0$, a quantum physicist is hardly different from a ``unary'' classical physicist.

The question ``Is the cat dead?'' is embedded in a \emph{field} of questions: we can ask ``Is the cat's heart working?'', ``Is the cat's liver working?'', ``has a particular hair on the cat's left rear leg fallen out?'', and so on and on, but CM$_+$ allows us also to ask differently integrative questions, such as ``Is the cat between alive and dead, with its heart working but with its liver not working?''
As we ask more and more detailed questions, and discover answers, we construct larger and larger Hilbert spaces that contain the Hilbert spaces we construct if we ask fewer questions (for quantum field theory, and for the equivalent mathematics in CM$_+$, every detailed question has the possibility of more detailed questions, ...)
Perhaps most importantly, experimental apparatus we can actually construct allows us to ask \emph{some} of the possible integrative questions ``in hardware'', not just by analysis of more detailed questions: we can construct diffraction gratings to answer questions about the weighted average of the answers to many detailed questions without knowing the answers to any of the individual questions.

Insofar as resuscitation of a long--dead cat is in practice impossible for either a classical or a quantum physicist, of course no--one can prepare an eigenstate of the Lewis Carroll operators.
If we can physically implement such reversals for a given real system, however, which in practice for some we can\cite{Frowis}, it can equally be modeled by a classical or a quantum physicist, and such operations can be used as a computational resource.
As a last word ---though in CM$_+$ there is arguably no last word--- we have been wont to discuss ``quantum systems'' as distinct from ``classical systems'', according to whether we can implement multiple clearly incompatible measurements, but it seems better to discuss ``nontrivial systems of measurements'', which is more neutral both as to the distinction between CM$_+$ and QM, and also as to whether there is such a thing as a distinguishable ``system''.


\section{Discussion}\label{Discussion}
We have here constructed a presentation of unary classical mechanics for which only the measurement theory is the same as a reasonable, if rather minimal, measurement theory for quantum mechanics, a relationship between states--and--operators and experimental raw data, which provides a conceptual bridge between unary classical mechanics and quantum mechanics.
We have deliberately made no attempt, however, to construct a more definite mathematical link between them, of a formal quantization procedure.
The informal connecting link is that for both there are statistical states over $*$--algebras, or, more simply, that for both there are \emph{``Hilbert spaces''} as a way to describe different experimental contexts and different signal analysis algorithms, but with different states over different $*$--algebras.
The change of perspective suggested here is almost no change at all: perhaps it might be as well to rename quantum computing, say, as ``Hilbert computing'', to demystify it, but the Hilbert space heart of the work is no different, except for, perhaps, a clearer understanding of measurement by analogy with classical measurement and signal analysis.

Although we can interpret unary classical mechanics in whatever way we interpret quantum mechanics, there is a significant difference in that for unary classical mechanics Planck's constant plays no part.
In a more elaborate framework of random fields and quantum fields, however, \emph{isomorphisms} can be constructed for some cases, including the physically important case of the electromagnetic field\cite{MorganEM,MorganEPL}, for which a clear symmetry group distinction can be seen between Poincar\'e invariant quantum fluctuations, with an \emph{action} scale determined by Planck's constant, and thermal fluctuations that are invariant under only the little group of the Poincar\'e group that is defined by a Hamiltonian operator, with an \emph{energy} scale determined by temperature and the Boltzmann constant\cite{MorganFluctuations}.
Such a distinction is of course not available in the absence of the 1+$n$--signature metric of Minkowski space.
The algebraic connection between the constructions given here for unary classical mechanics and quantum mechanics is nonetheless very close, in that Eq. (\ref{GeneratingFunction}) for the Gibbs equilibrium state of a classical simple harmonic oscillator is equally satisfied for the ground state of a quantized simple harmonic oscillator or the vacuum state of a free quantum field if the pre--inner product $({\bf f},{\bf g})$ is suitably replaced.
For the quantized electromagnetic field, in a manifestly Poincar\'e invariant construction\cite{MorganEM}, we have, as for Eq. (\ref{GeneratingFunction}),
\[\langle 0|\rme^{\,\rmj\lambda_1\hat\phi_{\!f_1}}
     \cdots\rme^{\,\rmj\lambda_n\hat\phi_{\!f_n}}|0\rangle
         =\exp\Big[\!-\!\Big(\sum_{i=1}^n\!\lambda_i f_{\!i}^{*\!}\!{,}\sum_{j=1}^n\!\lambda_{j\!}f_{\!j}\!\Big)/2\,{-}\!\sum\limits_{i<j}[(f_{\!i}^{*\!}\!,f_{\!j})\,{-}\,(f_{\!j}^{*\!}\!,f_{\!i})]/2\Big],
\]
\begin{equation}
(f,g) = -\hbar\!\!\int\!\!k^\alpha{{\tilde f}_{\alpha\mu}\strut}^{\hspace{-1.6ex}*
                          \hspace{0.6ex}}(k)\,\mathsf{g}^{\mu\nu}\,
                           k^\beta\tilde g_{\beta\nu}(k)
                         2\pi\delta(k{\cdot}k)\theta(k_0)\frac{\rmd^4k}{(2\pi)^4},
\end{equation}
with the metric tensor $\mathsf{g}^{\mu\nu}$ being constant of signature (1,-1,-1,-1) and 
$k^\alpha{{\tilde f}_{\alpha\mu}\strut}^{\hspace{-1.6ex}*\hspace{0.6ex}}(k)$ and
$k^\beta\tilde g_{\beta\nu}(k)$ are both space--like 4--vectors orthogonal to the light--like 4--vector $k$.
The algebraic structure is identical, the Weyl--Heisenberg group, but with different geometric structures.

There is a second difference, noted earlier, that although the Hamiltonian function of classical mechanics is positive the Liouvillian operator that generates evolution over time is not a positive operator, in contrast to the Hamiltonian operator that generates evolution over time in quantum theory, which \emph{is} positive.
This closely parallels the observation that the systematic use of quantum non-demolition measurement operators within the quantum mechanics formalism results in a generator of evolution over time that is non--positive\cite[Eq. (12)]{TsangCaves}.
This difference changes analytic properties of the dynamics significantly, however it does not change the abstract relationships between measurements and operator algebras and between statistics of measurement results and states.

The traditional connection between classical probabilities generated by theoretical physics and statistics of experimental raw data does not require a ``collapse'' of a state: the state reports probabilities, which are \emph{somehow} related to statistics of ensembles.
We here suggest a pragmatic case-by-case agnosticism about that relationship, without any stipulation that it must be Bayesian, frequentist, parameter estimation, or otherwise, although for those who have a strongly held adherence to a particular metaphysical interpretation of quantum mechanics, the Hilbert space mathematics for unary classical mechanics given here can be interpreted in that same way.
As was shown in \S\ref{MeasurementProblem}, ``collapse'' of the state is also not required insofar as we can insist on the classically natural Principle (JM), that joint measurements must be modeled by mutually commuting operators.
Thirdly, as noted in \S\ref{BellInequalities}, the electronic signals within an apparatus are not essentially discontinuous when they are considered at sufficient resolution: it is only when signals satisfy elaborate trigger conditions that we identify events.
It should also be noted that Gibbsian states are subject to question from a classical perspective\cite{FriggWerndl}.
Whatever interpretation one adopts, however, one cannot quite as easily say, for example, that at small scales the world is quantum, at large scales the world is classical, insofar as the classical is quantum too.
If we understand the measurement problem for classical mechanics, then we as much understand it for quantum mechanics.\vspace{3ex}

\section*{Acknowledgements}
  I am grateful to Gregor Weihs for access to the datasets used in \S\ref{BellInequalities}, to David Alan Edwards for a long sequence of incisive responses, to Jeremy Steeger for correspondence focused on the relationship between probability and statistics and experimental raw data, and to three anonymous reviewers.
Amongst many online comments and correspondence, I am especially grateful for comments that directly resulted in material changes, from Leslie Ballentine, Federico Comparsi, Richard Gill, Jean--Pierre Magnot, Ulla Mattfolk, and Arnold Neumaier.

\begin{appendix}
\section{Eq. (\ref{GeneratingFunction}) is a state}\label{GeneratingFunctionAppendix}
We can show directly for any operator $\hat A$ that $\rho(\hat A^\dagger\hat A)\ge 0$ (separately from the explicit construction of Eq. (\ref{GeneratingFunction})  as the usual ground state for an algebra of raising and lowering operators), the other properties required being straightforward.
$\hat A$ can be written as a sum of exponential terms, $\hat A=\sum_i\alpha_i\rme^{\,\rmj\hat F_{{\bf f}_i}}$, so that, applying Eq.  (\ref{GeneratingFunction}),
\[\rho(\hat A^\dagger\hat A)
       =\sum_{i,j}\alpha_i^*\alpha_j\rho(\rme^{\!-\,\rmj\hat F_{{\bf f}_i^*}}\rme^{\,\rmj\hat F_{{\bf f}_j}})
       =\sum_{i,j}\alpha_i^*\alpha_j\rme^{-({\bf f}_i,{\bf f}_i^*)/2-({\bf f}_j^*,{\bf f}_j)/2+({\bf f}_i,{\bf f}_j)}
       =\sum_{i,j}[\alpha_i\rme^{-({\bf f}_i^*,{\bf f}_i)/2}]^*
                                         \rme^{({\bf f}_i,{\bf f}_j)}[\alpha_j\rme^{-({\bf f}_j^*,{\bf f}_j)/2}]\ge 0,
\]
which is positive semi-definite because $\rme^{({\bf f}_i,{\bf f}_j)}$ is a Hadamard exponential of $({\bf f}_i,{\bf f}_j)$, which is a positive semi-definite matrix because it is a Gram matrix.

\section{Joint measurement instruments}\label{JointMeasurement}
We give here a joint measurement instrument account that parallels the more abstract discussion in \S\ref{MeasurementProblem}.
Following the account and notation given by Ballentine\cite[\S 3.3]{BallentineFoundPhys}, we consider measurements $\hat A$ and $\hat B$ that have discrete degenerate eigenvalues $a_i$ and $b_j$,
$$\hat A|a_{i\lambda}\rangle=a_i|a_{i\lambda}\rangle,\quad
    \hat B|b_{j\mu}\rangle=b_i|b_{j\mu}\rangle
$$
(we will omit the degenerate eigenvector indices $\lambda$ and $\mu$ except where necessary.)
To implement these measurements, we introduce measurement instruments $A$ and $B$ that are initially in vector states $|A_0\rangle$ and $|B_0\rangle$ and unitary evolutions
\begin{eqnarray*}
  \hat U_{\!{}_A}|a_{i}\rangle\otimes|A_0\rangle&=&|a_{i}\rangle\otimes|A_i\rangle,\\
  \hat U_{\!{}_B}|b_{j}\rangle\otimes|B_0\rangle&=&|b_{j}\rangle\otimes|B_j\rangle.
\end{eqnarray*}
By linearity, for a general vector $|\psi\rangle$,
$$\hat U_{\!{}_A}|\psi\rangle\otimes|A_0\rangle=\sum_i\langle a_i|\psi\rangle\cdot|a_i\rangle\otimes|A_i\rangle,$$
and similarly for $\hat U_{\!{}_B}$.
We apply first $\hat U_{\!{}_A}$ and then $\hat U_{\!{}_B}$,
\[
  \hat U_{\!{}_B}\hat U_{\!{}_A}|\psi\rangle\otimes|A_0\rangle\otimes|B_0\rangle
    =\sum_j\sum_i\langle b_j|a_i\rangle\langle a_i|\psi\rangle
                                                        \cdot|b_j\rangle\otimes|A_i\rangle\otimes|B_j\rangle,
\]
from which, using the Born rule, we extract probabilities
\begin{eqnarray*}
  P(A=a_i\,|\psi)&=&|\langle a_i|\psi\rangle|^2\\
  P(A=a_i\ \&\ B=b_j\,|\psi)&=&|\langle b_j|a_i\rangle\,\langle a_i|\psi\rangle|^2.
\end{eqnarray*}
The probability of a measurement result $B=b_j$ given that a measurement $A$ has been made, but averaging over its measurement results, is
\begin{equation}
   P(B=b_j\,|\psi\mbox{ and }A\mbox{ measured})
            =\sum_i|\langle b_j|a_i\rangle\,\langle a_i|\psi\rangle|^2,\label{BwithAmeasured}
\end{equation}
which differs from the probability of a measurement result $B=b_j$ given that a measurement $A$ was never made,
$$P(B=b_j\,|\psi)=\Bigl|\sum_i\langle b_j|a_i\rangle\,\langle a_i|\psi\rangle\Bigr|^2
                        =|\langle b_j|\psi\rangle|^2,$$
by the omission of ``interference'' terms, unless $\hat A$ and $\hat B$ commute.
We can rewrite Eq. (\ref{BwithAmeasured}), using a projection operator associated with each eigenvalue $a_i$,
$\hat P_i=\sum_\lambda|a_{i\lambda}\rangle\langle a_{i\lambda}|$, as
\begin{equation}
P(B=b_j\ |\psi\mbox{ and }A\mbox{ measured})
            =\sum_i \langle b_j|\hat P_i|\psi\rangle\,\langle\psi|\hat P_i|b_j\rangle,
\end{equation}
which corresponds, comparably to Eq. (\ref{LudersOfMeasurement}), to either
\begin{itemize}
\item a L\"uders transformed measurement $\sum_i\hat P_i|b_j\rangle\langle b_j|\hat P_i$ in the state
$|\psi\rangle\langle\psi|$, or
\item a measurement $|b_j\rangle\langle b_j|$ in the L\"uders transformed state
$\sum_i\hat P_i|\psi\rangle\langle\psi|\hat P_i$,
\end{itemize}
so, following the algebra, either we can say that a measurement of $B$ after a measurement of $A$ is not in general the same as a measurement of $B$ alone, or we can say that the measurement of $A$ changed the state.
We can say either that both descriptions are equally acceptable, or we can insist that one or the other description is preferred for specific contexts.
The third way, suggested by Principle (JM), is to require all operators that are used to model joint measurements to commute, so that the L\"uders transformer has no effect on subsequent measurements.

\section{A matrix model for \S\ref{BellInequalities}}\label{MatrixRep}
We give an extremal $4{\times}4$ matrix model that has the algebraic structure given for the operators in \S\ref{BellInequalities}:
\begin{eqnarray*}
\hat \mathsf{a}= \left[ 
{\begin{array}{rrrr}
1 & 0 & 0 & 0 \\
0 & -1 & 0 & 0 \\
0 & 0 & 1 & 0 \\
0 & 0 & 0 & -1
\end{array}}
 \right]\!, \ \hat \mathsf{a}'=\left[ 
{\begin{array}{rrrr}
0 & 1 & 0 & 0 \\
1 & 0 & 0 & 0 \\
0 & 0 & 0 & 1 \\
0 & 0 & 1 & 0
\end{array}}
 \right]\!,\\
\\
\hat \mathsf{b}=\left[ 
{\begin{array}{rrrr}
1 & 0 & 0 & 0 \\
0 & 1 & 0 & 0 \\
0 & 0 & -1 & 0 \\
0 & 0 & 0 & -1
\end{array}}
 \right]\!, \ \hat \mathsf{b}'=\left[ 
{\begin{array}{rrrr}
0 & 0 & 1 & 0 \\
0 & 0 & 0 & 1 \\
1 & 0 & 0 & 0 \\
0 & 1 & 0 & 0
\end{array}}
 \right]\!,
\end{eqnarray*}\vspace{0.25ex}%
using which we obtain for $\hat\mathsf{C}=\hat\mathsf{a}\hat\mathsf{b}+\hat\mathsf{a}\hat\mathsf{b}'
                          +\hat\mathsf{a}'\hat\mathsf{b}'-\hat\mathsf{a}'\hat\mathsf{b}$,
$$
\hat \mathsf{C}= \left[ 
{\begin{array}{rrrr}
1 & -1 & 1 & 1 \\
-1 & -1 & 1 & -1 \\
1 & 1 & -1 & 1 \\
1 & -1 & 1 & 1
\end{array}}
 \right]\!,\
\hat \mathsf{C}^2=\left[ 
{\begin{array}{rrrr}
4 & 0 & 0 & 4 \\
0 & 4 & -4 & 0 \\
0 & -4 & 4 & 0 \\
4 & 0 & 0 & 4
\end{array}}
 \right]\!.
$$
For this extremal model for $\hat\mathsf{C}$, we have $\|\hat \mathsf{C}\|^2\,{=}\,\|\hat \mathsf{C}^2\|\,{=}\,8$, $\hat \mathsf{C}^3\,{=}\,8\hat \mathsf{C}$, $\mathsf{Tr}[\hat \mathsf{C}]\,{=}\,0$, and $\mathsf{Tr}[\hat \mathsf{C}^2]\,{=}\,16$.

We can use a density matrix $\hat\rho\,{=}\,\psi\psi^\dagger\,{=}\,\frac{1}{16}(\hat \mathsf{C}^2-2\!\sqrt{2}\hat \mathsf{C})$, where $\psi$ is a unit length eigenvector of $\hat \mathsf{C}$,
$$\psi=\frac{1}{\sqrt{4\sqrt{2}}}\left[ 
\hspace{-0.3em}{\begin{array}{r}
\sqrt{\sqrt{2}-1} \\
\sqrt{\sqrt{2}+1}\\
-\sqrt{\sqrt{2}+1}\\
\sqrt{\sqrt{2}-1}
\end{array}}
 \hspace{-0.3em}\right]\!,$$
to construct a state for which $\rho(\hat \mathsf{C})\,{=}\,\mathsf{Tr}[\hat \mathsf{C}\hat\rho]\,{=}\,-2\!\sqrt{2}$,\newline
\centerline{$\rho(\hat \mathsf{a})\,{=}\,\rho(\hat \mathsf{a}')\,{=}\,\rho(\hat \mathsf{b})\,{=}\,\rho(\hat \mathsf{b}')\,{=}\,0$, and}
\centerline{$\rho(\hat \mathsf{a}\hat \mathsf{b})\,{=}\,\rho(\hat \mathsf{a}\hat \mathsf{b}')\,{=}\,\rho(\hat \mathsf{a}'\hat \mathsf{b}')\,{=}\,-\Half\!\sqrt{2}$, $\rho(\hat \mathsf{a}'\hat \mathsf{b})\,{=}\,\Half\!\sqrt{2}$.}
[[\hspace{0.3em}{\small For a density matrix $\hat\rho\,{=}\,\frac{1}{16}(\hat \mathsf{C}^2+2\!\sqrt{2}\hat \mathsf{C})$, we obtain $\mathsf{Tr}[\hat \mathsf{C}\hat\rho]{=}2\!\sqrt{2}$,
$\rho(\hat \mathsf{a}\hat \mathsf{b}){=}\rho(\hat \mathsf{a}\hat \mathsf{b}'){=}\rho(\hat \mathsf{a}'\hat \mathsf{b}'){=}\Half\!\sqrt{2}$, $\rho(\hat \mathsf{a}'\hat \mathsf{b}){=}\,{-}\Half\!\sqrt{2}$.}\hspace{0.3em}]]

\end{appendix}


\begin{thebibliography}{99}
\bibitem{LandsmanAlgQM}
  N. P. Landsman, ``Algebraic Quantum Mechanics'', in D. Greenberger, K. Hentschel, and F. Weinert (Eds.), \emph{Compendium of quantum physics}, Springer, Berlin, 2009, pp. 6-10.
https://doi.org/10.1007/978-3-540-70626-7\_3

\bibitem{Haag}
  R. Haag, \emph{Local Quantum Physics: Fields, Particles, Algebras}, 2nd Edn., Springer, Berlin, 1996.

\bibitem{TsangCaves}
  M. Tsang, C. Caves,
``Evading Quantum Mechanics: Engineering a Classical Subsystem within a Quantum Environment'',
Phys. Rev. X \textbf{2} (2012) 031016. https://doi.org/10.1103/PhysRevX.2.031016

\bibitem{Koopman}
  B. O. Koopman,
``Hamiltonian Systems and Transformations in Hilbert Space'',
Proc. Natl. Acad. Sci. \textbf{17} (1931) 315.\\ https://doi.org/10.1073/pnas.17.5.315

\bibitem{Mauro}
  D. Mauro,
``A new quantization map'',
Phys. Lett. A \textbf{315} (2003) 28. https://doi.org/10.1016/S0375-9601(03)00996-4

\bibitem{Ghose}
  P. Ghose,
``The Unfinished Search for Wave--Particle and Classical--Quantum Harmony'',
J. Adv. Phys. \textbf{4} (2015) 236.\\ https://doi.org/10.1166/jap.2015.1197

\bibitem{MorganEM}
  P. Morgan,
``Classical states, quantum field measurement'',
Physica Scripta \textbf{94} (2019) 075003. https://doi.org/10.1088/1402-4896/ab0c53

\bibitem{Zalamea}
  F. Zalamea,
``The Twofold Role of Observables in Classical and Quantum Kinematics'',
Found. Phys. \textbf{48} (2018) 1061.\\ https://doi.org/10.1007/s10701-018-0194-8

\bibitem{BuchholzFredenhagen}
  D. Buchholz, K. Fredenhagen,
``Classical dynamics, arrow of time, and genesis of the Heisenberg commutation relations''
[arXiv:1905.02711].

\bibitem{Wetterich}
  C. Wetterich,
``Quantum formalism for classical statistics'',
Ann. Phys. \textbf{393} (2018) 1. https://doi.org/10.1016/j.aop.2018.03.022

\bibitem{tHooft}
  G. 't Hooft,
  \emph{The Cellular Automaton Interpretation of Quantum Mechanics},
    Springer, Cham, Switzerland, 2016.

\bibitem{Janotta}
  P. Janotta, H. Hinrichsen,
``Generalized probability theories: what determines the structure of quantum theory?'',
J. Phys. A \textbf{47} (2014) 323001. https://doi.org/10.1088/1751-8113/47/32/323001

\bibitem{Wang}
  S. Wang, H. I. Nurdin, G. Zhang, M. R. James,``Representation and network synthesis for a class of mixed
quantum--classical linear stochastic systems'',
  Automatica \textbf{96} (2018) 84. https://doi.org/10.1016/j.automatica.2018.06.003

\bibitem{James}
  H. R. James, H. I. Nurdin, I. R. Petersen,``$H^\infty$ Control of Linear Quantum Stochastic Systems'',
  IEEE Transactions on Automatic Control \textbf{53} (2008) 1787. https://doi.org/10.1109/TAC.2008.929378

\bibitem{Gough}
  J. Gough, M. R. James, ``The Series Product and Its Application to Quantum Feedforward and Feedback Networks'',
  IEEE Transactions on Automatic Control \textbf{54} (2009) 2530. https://doi.org/10.1109/TAC.2009.2031205

\bibitem{LeonCohen}
  L. Cohen,
``Time--Frequency Distributions --- A Review'',
Proc. IEEE \textbf{77} (1989) 941. https://doi.org/10.1109/5.30749

\bibitem{Kisil}
  V. Kisil,
``Symmetry, geometry and quantization with hypercomplex numbers'',
Geometry, Integrability and Quantization \textbf{18} (2017) 11.\\ http://dx.doi.org/10.7546/giq-18-2017-11-76

\bibitem{3Blue1Brown}
  3Blue1Brown, ``The more general uncertainty principle, beyond quantum'',\\ https://www.youtube.com/watch?v=MBnnXbOM5S4 (accessed 8 June 2019).
\bibitem{MinutePhysics}
  Minute Physics, ``What is the Uncertainty Principle?'',\\
https://www.youtube.com/watch?v=7vc-Uvp3vwg (accessed 8 June 2019).
\bibitem{SixtySymbols}
  Sixty Symbols, ``The Uncertainty Principle and Waves'',\\ https://www.youtube.com/watch?v=VwGyqJMPmvE (accessed 8 June 2019).
\bibitem{ScienceAsylum}
  The Science Asylum, ``Wave--Particle Duality and other Quantum Myths'',\\ https://www.youtube.com/watch?v=Q2OlsMblugo (accessed 8 June 2019).

\bibitem{WeinbubFerry}
  J. Weinbub, D. K. Ferry,
``Recent advances in Wigner function approaches'',
Appl. Phys. Rev. \textbf{5} (2018) 041104.\\ https://doi.org/10.1063/1.5046663

\bibitem{Ballentine}
  L. E. Ballentine, \emph{Quantum Mechanics: A Modern Development}, World Scientific, Singapore, 1998.

\bibitem{Woodhouse}
  N. M. J. Woodhouse, \emph{Geometric Quantization}, 2nd Edn., Oxford University Press, Oxford, 1997.

\bibitem{JackCohn}
  J. Cohn,
``Operator formulation of classical mechanics'',
Am. J. Phys. \textbf{48} (1980) 379. https://doi.org/10.1119/1.12109

\bibitem{Belavkin}
  V. P. Belavkin,
``Nondemolition Principle of Quantum Measurement Theory'',
Found. Phys. \textbf{24} (1994) 685.\\ https://doi.org/10.1007/BF02054669

\bibitem{CohenCF}
  L. Cohen,
``Rules of Probability in Quantum Mechanics'',
Found. Phys. \textbf{18} (1988) 983. https://doi.org/10.1007/BF01909934

\bibitem{Moretti}
  V. Moretti, \emph{Spectral Theory and Quantum Mechanics: Mathematical Foundations of Quantum
Theories, Symmetries and Introduction to the Algebraic Formulation}, 2nd Edn., Springer, Cham, Switzerland, 2017.

\bibitem{Landsman}
  K. Landsman, \emph{Foundations of Quantum Theory: From Classical Concepts to Operator Algebras}, Springer, Cham, Switzerland, 2017.

\bibitem{Steeger}
  J. Steeger,
``Probabilism for stochastic theories'',
Studies in History and Philosophy of Modern Physics \textbf{66} (2019) 34.\\ https://dx.doi.org/10.1016/j.shpsb.2018.10.004

\bibitem{Chang}
  H. Chang, \emph{Inventing Temperature: Measurement and Scientific Progress}, Oxford University Press, Oxford, 2004.

\bibitem{Landau}
  L. J. Landau,
``On the violation of Bell's inequality in quantum theory'',
Phys. Lett. A \textbf{120} (1987) 54.\\
https://doi.org/10.1016/0375-9601(87)90075-2

\bibitem{deMuynck}
  W. M. de Muynck,
``The Bell inequalities and their irrelevance to the problem of locality in quantum mechanics'',
Phys. Lett. A \textbf{114} (1986) 65. https://doi.org/10.1016/0375-9601(86)90480-9

\bibitem{Griffiths}
  R. B. Griffiths,
``Quantum Nonlocality: Myth and Reality''
[arXiv:1901.07050] \S 2.

\bibitem{AcharyaKypraiosGuta}
  A. Acharya, T. Kypraios, M. Gu\c t\v a,
``A comparative study of estimation methods in quantum tomography'',
J. Phys. A \textbf{52} (2019) 234001.\\ https://doi.org/10.1088/1751-8121/ab1958

\bibitem{BuschGrabowskiLahti}
  P. Busch, M. Grabowski, P. J. Lahti, \emph{Operational Quantum Physics}, Springer, Berlin, 1997.

\bibitem{Shankar}
  R. Shankar, \emph{Principles of quantum mechanics}, 2nd Edn., Springer, New York, 1994.

\bibitem{Mott}
  N. F. Mott,
``The wave mechanics of $\alpha$--Ray tracks'', Proc. R. Soc. Lond. A \textbf{126} (1929) 79.
http://doi.org/10.1098/rspa.1929.0205

\bibitem{LiangSpekkensWiseman}
  Y.--C. Liang, R. W. Spekkens, H. M. Wiseman,
``Specker's parable of the overprotective seer: A road to contextuality, nonlocality and complementarity'',
Physics Reports \textbf{506} (2011) 1. https://doi.org/10.1016/j.physrep.2011.05.001

\bibitem{Ottinger}
  H. C. \"Ottinger, \emph{A Philosophical Approach to Quantum Field Theory}, Cambridge University Press, Cambridge, 2018.

\bibitem{Howard}
  D. Howard, ``Who Invented the “Copenhagen Interpretation”? A Study in Mythology'', Philosophy of Science \textbf{71} (2004) 669. https://doi.org/10.1086/425941

\bibitem{Oldofredi}
  A. Oldofredi, M. Esfeld, ``Observability, Unobservability and the Copenhagen Interpretation in Dirac's Methodology of Physics'', Quanta \textbf{8}, (2019) 68. https://doi.org/10.12743/quanta.v8i1.93



\bibitem{BellReview}
  N. Brunner, D. Cavalcanti, S. Pironio, V. Scarani, S. Wehner,
``Bell nonlocality'',
Rev. Mod. Phys. \textbf{86} (2014) 419.\\ https://doi.org/10.1103/RevModPhys.86.419

\bibitem{Weihs}
  G. Weihs, T. Jennewein, C. Simon, H. Weinfurter, A. Zeilinger,
``Violation of Bell’s Inequality under Strict Einstein Locality Conditions'',
Phys. Rev. Lett. \textbf{81} (1998) 5039. https://doi.org/10.1103/PhysRevLett.81.5039

\bibitem{WeihsThesis}
  G. Weihs, ``Ein Experiment Zum Test Der Bellschen Ungleichung Unter Einsteinscher Lokalit\"at'', (Ph.D. Thesis),
University of Vienna, 2000. http://www.uibk.ac.at/exphys/photonik/people/gwdiss.pdf

\bibitem{MorganBellForRandom}
  P. Morgan, ``Bell inequalities for random fields'',
J. Phys. A \textbf{39} (2006) 7441. https://doi.org/10.1088/0305-4470/39/23/018

\bibitem{Frowis}
F. Fr\"owis, P. Sekatski, W. D\"ur, N. Gisin, N. Sangouard,
``Macroscopic quantum states: Measures, fragility, and implementations'',
Rev. Mod. Phys. \textbf{90} (2018) 025004. https://doi.org/10.1103/RevModPhys.90.025004

\bibitem{Serafini}
  A. Serafini, M. G. A. Paris, F. Illuminati, S. De Siena,
``Quantifying decoherence in continuous variable systems'',
J. Opt. B \textbf{7} (2005) R19. https://doi.org/10.1088/1464-4266/7/4/R01

\bibitem{MorganEPL}
  P. Morgan, ``Equivalence of the Klein--Gordon random field and the complex Klein--Gordon quantum field'',
EPL \textbf{87} (2009) 31002.\\ https://doi.org/10.1209/0295-5075/87/31002

\bibitem{MorganFluctuations}
  P. Morgan, ``A succinct presentation of the quantized Klein--Gordon field, and a similar quantum presentation of the classical Klein--Gordon random field'',
Phys. Lett. A \textbf{338} (2005) 8. https://doi.org/

\bibitem{FriggWerndl}
  R. Frigg, C Werndl, ``Can Somebody Please Say What Gibbsian Statistical Mechanics Says?'',
Brit. J. Phil. Sci. (to be published).\\ https://doi.org/10.1093/bjps/axy057

\bibitem{BallentineFoundPhys}
  L. E. Ballentine,
``Limitations of the Projection Postulate'',
Found. Phys. \textbf{20} (1990) 1329. https://doi.org/10.1007/BF01883489



\end{thebibliography}
\end{document}